\documentclass[preprint2,times,trackchanges]{aastex631}
\usepackage{amssymb, amsmath, amsthm, multirow, makecell}
\usepackage[table]{xcolor}

\shorttitle{Radial Profiles of Large-scale Filaments}
\shortauthors{Su, Wang et al.}

\graphicspath{{./}}

\begin{document}

\title{Galactic Large-scale Filaments Resident in Asymmetric Environments: \\
Clues from Cross-filament Profiles of Density and Temperature}

\correspondingauthor{Ke Wang}
\email{kwang.astro@pku.edu.cn}

\author[0009-0003-2243-7983]{Keyun Su}
\affiliation{Kavli Institute for Astronomy and Astrophysics, Peking University, Haidian District, Beijing 100871, People’s Republic of China}
\affiliation{Department of Astronomy, School of Physics, Peking University, Beijing, 100871, People’s Republic of China}

\author[0000-0002-7237-3856]{Ke Wang}
\affiliation{Kavli Institute for Astronomy and Astrophysics, Peking University, Haidian District, Beijing 100871, People’s Republic of China}

\author[0000-0001-5950-1932]{Fengwei Xu}
\affiliation{Kavli Institute for Astronomy and Astrophysics, Peking University, Haidian District, Beijing 100871, People’s Republic of China}
\affiliation{Max Planck Institute for Astronomy, Königstuhl 17, 69117 Heidelberg, Germany}

\author[0000-0001-8812-8460]{N. K. Bhadari}
\affiliation{Kavli Institute for Astronomy and Astrophysics, Peking University, Haidian District, Beijing 100871, People’s Republic of China}

\begin{abstract}
Large-scale filaments ubiquitously exist in the Galactic interstellar medium, and their radial profiles offer insights into their formation mechanisms. We present a statistical analysis of molecular hydrogen column density ($\rm N(H_2)$) and dust temperature ($\rm T_d$) radial profiles for 35 Galactic large-scale filaments. We divided their spines into 315 segments, extracted the radial profiles of each segment using $\rm N(H_2)$ and $\rm T_d$ maps derived from $Herschel$ Hi-GAL data, and estimated the asymmetry degree within the radial profiles ($\alpha_{\rm asy}$), as well as the length proportion of segments with asymmetric profiles across the entire filament ($f_{\rm asy}$). 
We found that Galactic large-scale filaments reside in surroundings distinctly asymmetric and varied in $\rm N(H_2)$, and mild asymmetric yet stable in $\rm T_d$.
Different filament morphology types do not show significant differences in $\alpha_{\rm asy}$ or $f_{\rm asy}$. A bent filament shape does not necessarily correspond to an asymmetric radial profile, whereas a straight filament shape may be associated with a symmetric profile.
Segments with asymmetric surroundings in $\rm N(H_2)$ may not simultaneously appear asymmetric in $\rm T_d$, and vice versa.
We found three filaments with 4-44\% of their spine show asymmetric $\rm N(H_2)$ and $\rm T_d$ radial profiles in inverse trends, likely caused by nearby H\textsc{ii} region. H\textsc{ii} regions of similar scale to large filaments can induce asymmetric radial profiles within them, indicating their influence on filament evolution. However, they are unlikely to independently trigger the formation of an entire Galactic large-scale filament, in contrast to their role in small-scale filament formation.
\end{abstract}


\section{Introduction} \label{sec:intro}
Galactic filaments are elongated structures in molecular clouds that are overdense and skinny compared to their surrounding environments \citep{Andre2014}. They were first discovered as dark clouds in infrared Galactic surveys \citep[e.g.,][]{Schneider1979, McClure2006}, and later the Herschel Space Observatory has revealed their ubiquitous presence in the Galactic interstellar medium \citep[][]{Andre2010, Molinari2010}. 
Velocity coherence has verified that filaments, from large to small, are physical entities rather than chance alignments. They can play an important role in star formation, and their sizes range from less than 1 pc to more than 100 pc \citep[e.g., ][]{Andre2014, Hacar2023}. Among these, filaments longer than 10 pc (also known as Giant Molecular Filaments) have been of special interest \citep{Wang2015, Wang2016, Ge2022, Ge2023, Wang2024}. With this length scale, they can be influenced by large-scale effects including Galactic shear and superbubbles at tens to hundreds of pc \citep{Clarke2023}, as well as experiencing feedback from star-forming regions on pc or sub-pc scales. Although current simulations have managed to produce filamentary structures through various combinations of gravity, magnetic fields, and turbulence \citep[e.g.,][]{Inutsuka2015, Zucker2019}, the varying settings of these simulations mean that we are still unable to reach a consensus on their formation processes \citep{Pineda2023}.

Although complicated, some filament formation mechanisms can be relatively easily identified through observations. One such mechanism involves material redistribution triggered by H\textsc{ii} regions on 1-10 pc scale. Their expansion causes the interstellar medium to accumulate at the rim of their shell, generating a compressed layer that subsequently undergoes fragmentation into molecular clouds, filaments and dense clumps \citep{Inutsuka2015, Dirienzo2012, Zhang2021, Pineda2023, Zhang2023, Xu2024}. From an observational perspective, the expansion of an H\textsc{ii} region pushes the once uniformly distributed material outward, causing the region inside the shell to become less dense than the outside, resulting in an asymmetric density radial profile. Simultaneously, heating from the central star raises the temperature inside the H\textsc{ii} region, generating an asymmetric temperature radial profile \citep{Inutsuka2015, Pineda2023}. Such asymmetric features in radial profiles can serve as tracers to identify whether an H\textsc{ii} region is influencing the filament formation process. For example, \citet{Zavagno2020} extracted the radial profiles of two small pc-scale filaments surrounding the central H\textsc{ii} region in the RCW120 system. They found that both filaments exhibited asymmetric features in their 350 $\mu \rm m$ flux intensity radial profiles, with the side closer to the center of the H\textsc{ii} region having a lower flux intensity, suggesting that the central H\textsc{ii} region likely contributed to their formation. 

However, for large-scale filaments (LSF), this theory has not yet been systematically tested. \cite{Wang2015} first studied the cross-filament profiles of 9 prominent $Herschel$-identified LSFs and reported the asymmetric molecular hydrogen column density (hereafter $\rm N(H_2)$) and dust temperature (hereafter $\rm T_d$) radial profiles in the $>$ 50 pc-long filament G64. \citet{Clarke2023} identified an asymmetric $\rm N(H_2)$ radial profile in the 35-pc-long filament G214.5-1.8. Combined with an H\textsc{i} Position-Velocity (PV) slice, they suggested that the filament might be compressed by a nearby H\textsc{i} superbubble. To determine whether asymmetric features are common in the radial profiles of Galactic LSFs, whether these asymmetric profiles in LSFs are generated by H\textsc{ii} regions like in their small-scale counterparts, and how much impact H\textsc{ii} regions can bring to the formation of Galactic LSFs, a systematic search in a relatively large sample of Galactic LSFs is still needed.

Developments in telescope sensitivity and filament identification algorithms have enabled the discovery of a large number of Galactic LSF candidates under uniform selection criteria. \citet{Wang2016} introduced the Minimum Spanning Tree (MST) algorithm as an efficient and less biased method to identify Galactic LSFs. The concept of MST originates in graph theory, its goal is to use a combination of segments to connect a given set of points, with the requirements of segments not forming closed loops and ensuring their total length is minimized \citep{Prim1957}. In the field of star formation, the MST algorithm is widely used for identifying star clusters, particularly for estimating the separations between components within star clusters \citep{Battinelli1991, Hetem1993, Cartwright2004, Gutermuth2009, Kirk2014}. When applied to LSF identification, the methodology is similar: the MST algorithm is used to generate a set of segments to connect dense star-forming clumps selected from existing catalogs. The connected clumps and their linking segments together construct the MST estimation of an LSF candidate. Each pair of connected clumps should not only be spatially close but also exhibit similar line-of-sight velocities to maintain the coherency of the LSF candidate. By applying the MST algorithm to the Bolocam Galactic Plane Survey clump catalog \citep[BGPS;][]{Rosolowsky2010}, \citet{Wang2016} identified a total of 56 Galactic LSFs across the sky range of $7^{\circ}.5\ \leq\ l\ \leq\ 194^{\circ}$. As a follow-up, \citet{Ge2022} applied the MST algorithm to the APEX Telescope Large Area Survey of the Galaxy clump catalog \citep[ATLASGAL;][]{Urquhart2014}, identifying 163 Galactic LSFs within the sky range of $|l|\ \leq\ 60^{\circ}$. Together, these studies have provided a sample of over 200 Galactic LSFs, sufficient to conduct a systematic search for asymmetric radial profiles within them.

At the same time, high-resolution surveys at radio wavelengths and existing H\textsc{ii} region catalogs can help identify the locations of H\textsc{ii} regions, enabling an examination of their relationship with asymmetric radial profiles in Galactic LSFs. The MeerKAT survey at radio wavelengths \citep{Goedhart2024} and the all-sky Wide-Field Infrared Survey Explorer \citep[WISE, ][]{Anderson2014} H\textsc{ii} region catalog \citep{Anderson2014} have facilitated multi-wavelength identification of H\textsc{ii} regions within the Galactic plane. Combined with the numerous Galactic LSF candidates, it is now possible to select a set of LSFs, extract their radial profiles, identify asymmetric features, and examine their association with nearby H\textsc{ii} regions. This approach allows for a systematic test of the H\textsc{ii} region-triggered filament formation scenario in Galactic LSFs, which can potentially provide insights into filament formation mechanisms and offer observational constraints for numerical simulations of Galactic LSFs.

This paper targets filament samples from \citet{Wang2015}, \citet{Wang2016}, \citet{Ge2022} and studies the asymmetry in their radial profiles of $\rm N(H_2)$ and $\rm T_d$. The structure of this paper is as follows: the data used to extract radial profiles and the filament selection criteria are described in Section \ref{sec:sourcedata}. Methods for extracting radial profiles and estimating asymmetric profiles are presented in Section \ref{sec:methodresults}. Further discussion on the relationship between asymmetric profiles and nearby H\textsc{ii} regions follows in Section \ref{sec:discussion}, and conclusions are summarized in Section \ref{sec:conclusion}.

\section{Data} \label{sec:sourcedata}
\subsection{Column Density and Temperature Maps} \label{subsec:HiGAL}
We use the $\rm N(H_2)$ maps and $\rm T_d$ maps constructed by \citet{Marsh2017} to extract radial profiles. These maps are generated by applying the Point Process Mapping \citep[PPMAP;][]{Marsh2015} algorithm to the Hi-GAL multi-wavelength continuum maps \citep{Molinari2010}.

Hi-GAL is an unbiased photometric survey at far-IR wavelengths carried out with the $Herschel$ satellite \citep{Pilbratt2010}. The survey covers the entire Galactic plane with a Galactic latitude range of $|b| \leq 1^\circ$. Its dataset includes $2^\circ \times 2^\circ$ dust emission maps at 70, 160, 250, 350, and 500 $\mu \rm m$ \citep{Molinari2010}. PPMAP is an algorithm designed to convert dust emission maps into multiple column density maps under given dust temperatures, assuming the dust is optically thin \citep{Marsh2015}. By applying PPMAP to the Hi-GAL dust emission maps, \citet{Marsh2017} produced 163 data cubes to depict the $\rm N(H_2)$ distribution in different $\rm T_d$ slices across the entire Galactic plane. Each data cube covers a sky region of $2^{\circ}.4 \times 2^{\circ}.4$ with an angular resolution of 12\arcsec. The temperature axis consists of 12 logarithmically spaced $\rm T_d$s ranging from 8 K to 50 K. Based on these data cubes, \citet{Marsh2017} further calculated two-dimensional temperature-integrated $\rm N(H_2)$ maps and density-weighted mean $\rm T_d$ maps. We used these two sets of maps to extract the density and temperature radial profiles of the LSFs.

In addition to the data cubes of $\rm N(H_2)$ across different temperature slices, \citet{Marsh2017} also provided corresponding uncertainty data, stored in cubes with the same sky coverage, resolution and temperature slices. Based on these uncertainty cubes and the method described in \citet{Marsh2017} for generating the two-dimensional $\rm N(H_2)$ and $\rm T_d$ maps, we calculate the uncertainty maps of the temperature-integrated $\rm N(H_2)$ and density-weighted mean $\rm T_d$ according to the propagation of uncertainty. The detailed formulas for these calculations are provided in Appendix \ref{App_ErrorProp}. These uncertainty maps are used to derive the uncertainties of the extracted density and temperature radial profiles.

\subsection{Datasets for H\textsc{ii} region identification} \label{subsec:HIIdata}
We use a combination of spectral index maps from the SARAO MeerKAT 1.3 GHz Galactic Plane Survey \citep[SMGPS, ][]{Goedhart2024} and the WISE catalog \citep{Anderson2014} to identify H\textsc{ii} regions around the LSFs. 

The SMGPS is a radio-band survey observing continuum emission at approximately 1.3 GHz, covering sky ranges of $51^{\circ} \leq l \leq 358^{\circ}$, $2^{\circ} \leq l \leq 61^{\circ}$, with $|b| \leq 1.5^{\circ}$. It offers an angular resolution of 8\arcsec and a sensitivity of approximately $10$-$20\ \rm \mu Jy\ beam^{-1}$. Its first data release includes emission map data cubes in 14 frequency channels, pixel-by-pixel SED fitting results, the derived spectral index maps, and zeroth-moment maps averaged across the observed bands \citep{Goedhart2024}.

The WISE catalog contains the positions of 8399 H\textsc{ii} regions across the entire Galactic plane with $|b| \leq 8^{\circ}$. H\textsc{ii} regions in the WISE catalog are identified through visual inspection of circular-like structures in the WISE 12 $\mu \rm m$ and 22 $\mu \rm m$ emission maps, as well as in radio continuum images \citep[see Table 1 of ][]{Anderson2014}. Based on the availability of corresponding radio recombination line or H$\alpha$ data, positional association with known H\textsc{ii} region complexes, and the presence of radio emission, they are categorized into five types: ``K" (known), ``G" (group), ``C" (candidate), ``Q" (radio-quiet), and ``?" (no radio data). A total of 1413 H\textsc{ii} regions in the catalog have distance estimates determined through maser parallax and kinematics methods \citep{Anderson2014}.

\subsection{Filament Sample} \label{subsec:filsample}
Our filament samples are mainly chosen from two existing Galactic LSF catalogs: \citet{Wang2016} and \citet{Ge2022}. These catalogs were constructed using the MST algorithm with the following criteria: (1) each LSF candidate should contain at least five clumps; (2) the length of each segment should be shorter than $0^{\circ}.1$; (3) each pair of connected clumps should have a line-of-sight velocity $v_{\rm lsr}$ difference less than 2 $\rm km\ s^{-1}$; (4) the total length of the filament spine should be greater than 10 pc; (5) the filament linearity $f_L$, defined as the ratio between the standard deviation of clump positions along the filament's major axis and that along the minor axis, should be greater than 1.5. Criteria (2) and (3) were designed to ensure the spatial and kinematic coherence of the LSF candidate. The specific numerical thresholds were established based on the MST algorithm's ability to reproduce the trajectory of the known LSF G11 \citep{Wang2016, Ge2022}. In criterion (5), the major and minor axes of the LSFs were obtained by fitting a straight line according to the (l,b) positions of the clumps included in the LSF candidate. For details of the fitting process we refer to Appendix B of \citet{Ge2022}; the code is publicly available at \citet{Wang2021_MST_code}. A larger $f_L$ indicates a more elongated distribution of clumps along the fitted major axis, implying a morphology closer to a linear filament. If an LSF candidate has $f_L\approx1$, the clumps within it will exhibit a circular-like distribution, contradicting the slender morphology of filaments. \citet{Wang2016} and \citet{Ge2022} adopted $f_L \geq 1.5$ to exclude false identifications.

After removing overlaps, the two catalogs present us a total of 175 unique LSF candidates. We apply the following steps to select the most prominent ones as part of our sample. First, we examine the distribution of the MST-estimated spines of the LSF candidates on the Hi-GAL $\rm N(H_2)$ map. For each LSF candidate, we visually inspect a pattern of connected segments that not only consistently trace positions with $\rm N(H_2)>3\ \sigma_{\rm N(H_2)}$, but also has a shape closest to the dense region in Hi-GAL $\rm N(H_2)$ map, while maximizing the total length. This connected set of segments is treated as the major branch, while other segments are considered as secondary branches. Next, we limit our samples to LSF candidates with fewer than five secondary branches. Extracting radial profiles requires sampling in the direction perpendicular to the filament spine, whereas secondary branches generally correspond to other dense structures around the filament. Their presence can complicate the radial profiles. Since LSF candidates identified by the MST algorithm often contain multiple secondary branches, setting a limit of five yields the simplest, no-branch filaments, or filaments with secondary branches negligible in length compared to the major branch. Finally, we assess velocity coherence along the major branch of each LSF candidate. We extract PV slices along the major branch using $\rm ^{13}CO$ spectral line emission data cubes from the Structure, Excitation and Dynamics of the Inner Galactic Interstellar Medium survey \citep[SEDIGISM, ][]{Schuller2021, Duarte-Cabral2021}, the FOREST Unbiased Galactic Plane Imaging Survey with the Nobeyama 45-m Telescope project \citep[FUGIN, ][]{Umemoto2017}, and the Galactic Ring Survey \citep[GRS, ][]{Jackson2006}. These steps result in a set of 29 LSF candidates, 24 from \citet{Ge2022} and 5 from \citet{Wang2016}, which we identify as the most prominent LSFs discovered via the MST algorithm and are thus included in our sample.

The MST-based LSF identification criteria in \citet{Ge2022} and \citet{Wang2016} are designed based on their effectiveness in reproducing the shape of G11. However, Galactic LSFs exhibit a wide range of evolutionary stages and environmental conditions, leading to diverse clump distributions within them. Considering the possibility that some LSFs with clump distribution different from G11 may not have been covered by \citet{Wang2016} and \citet{Ge2022}, we further include six LSFs: G26, G29, G47, G49, G64, and G339 from \citet{Wang2015}. These filaments are identified via visual inspection of strong emission in the 350 $\mu \rm m$ and 500 $\mu \rm m$ dust continuum images from the $Herschel$ Hi-GAL project \citep[Herschel Infrared Galactic plane survey;][]{Molinari2010}. Their velocity coherence is tested using PV slices generated from $\rm ^{13}CO$ data cubes from the GRS \citep{Jackson2006}. 

\begin{figure*}[!t]
    \centering
    \includegraphics[width = 1.\textwidth]{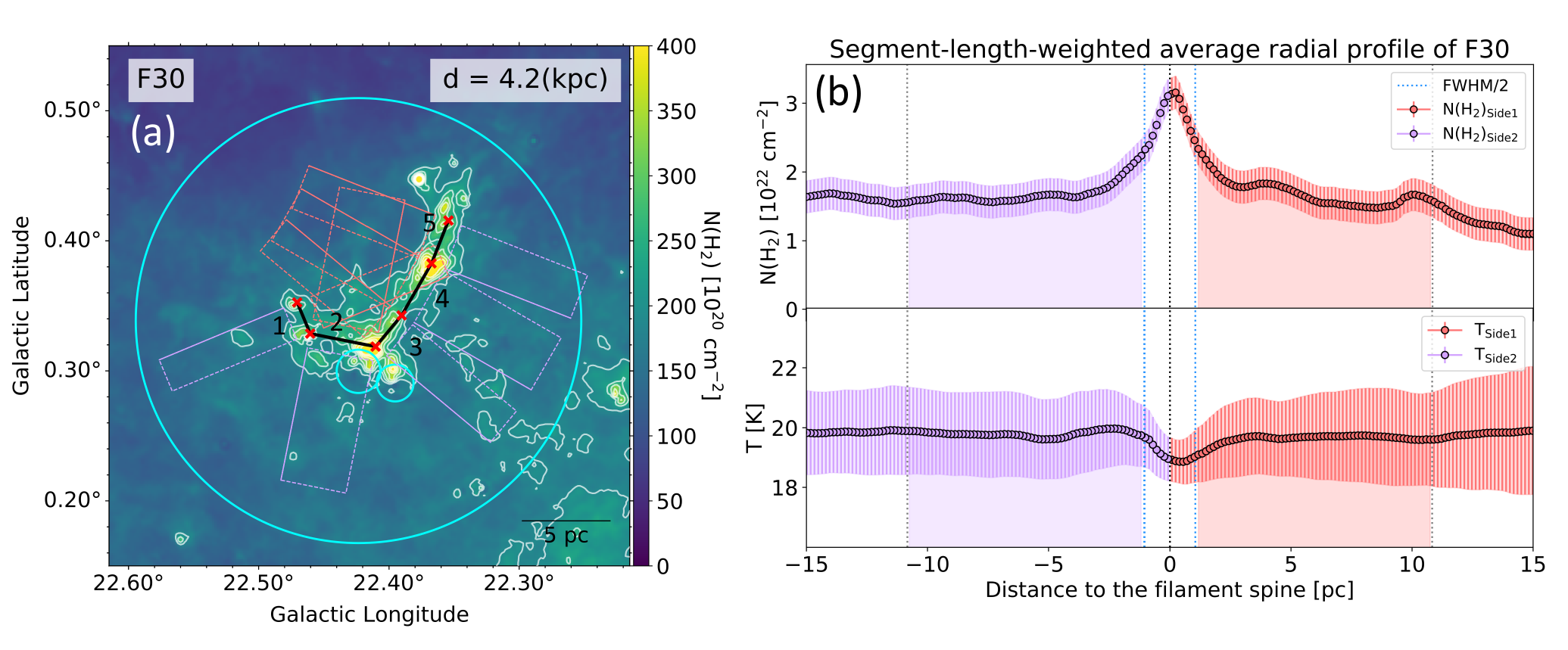}
    \caption{(a) Overview of the surroundings of filament sample F30 from \citet{Ge2022}. The background is the Hi-GAL-based $\rm N(H_2)$ map derived by \citet{Marsh2017}, with contour levels ranging from 4 $\sigma_{\rm N(H_2)}$ to 10 $\sigma_{\rm N(H_2)}$ ($\sigma_{\rm N(H_2)} = 47.52\times 10^{20}\ \rm cm^{-2}$). The distance to the filament is indicated in the upper-right corner, and a 5 pc scale bar is shown in the lower-right corner of the image. The MST spine of the filament is represented by red crosses connected by black line segments, with the ID for each segment labeled around them. The boxes perpendicular to the filament spine frame the areas used to extract radial profiles for each segment. Their shapes imitate the ``projection" regions in DS9, with the solid line representing the direction of extracting radial profiles. The lengths of the boxes are calculated in Section \ref{subsec:radprofmethod}. The cyan circles correspond to nearby H\textsc{ii} regions, adopted from the WISE catalog \citep{Anderson2014}, with their distances in kpc labeled nearby, if available. (b) $\rm N(H_2)$ and $\rm T_d$ radial profiles of F30. The $\rm N(H_2)$ profile is shown in the upper panel, and the $\rm T_d$ profile in the lower panel. The blue vertical dotted line represents the start of the sampling region as half of the filament width. Regions shaded red and purple correspond to radial profiles extracted from the red and purple sampling regions in (a). Only these parts are taken into subsequent calculations and identification of asymmetry in the LSF. } 
    \label{fig1_F30}
\end{figure*}

The properties of the 35 selected Galactic LSFs are listed in Table \ref{tab:35fils}. Since G26 is a bifurcated filament with two branches of significant length, we treat each branch as a separate filament sample, labeling them as ``G26L" and ``G26U", where L denotes the lower branch and U denotes the upper branch. Distances of the MST-based LSFs are defined as the median distance of the clumps connected within the MST LSF spine \citep{Wang2016, Ge2022}. \citet{Ge2022} estimated a universal distance uncertainty of 0.3 kpc for their LSFs. For LSFs from \citet{Wang2016}, we use the standard deviation of the distances of the connected clumps to represent their distance uncertainty. LSFs from \citet{Wang2015} have kinematic distances estimated based on the spiral arm models constructed by \citet{Reid2009} and \citet{Reid2014}. The widths of the LSFs are defined as the Full Width at the Half Maximum (FWHM) of Plummer-like $p=4$ fits to their segment-length-weighted average Hi-GAL $\rm N(H_2)$ radial profile \citep{Suri2019}. We note that this fitting method may not be optimal for all LSFs, as their environments and evolutionary stages differ, and the degree of smoothing varies with distance \citep{Pineda2011, Hacar2018, Schmiedeke2021}. Therefore, the LSF widths derived through this procedure should be regarded as reference values, used primarily to establish a uniform criterion for defining the sampling length. We adopt the morphology classification result of each LSF from their parent catalogs. \citet{Wang2015}, \citet{Wang2016}, and \citet{Ge2022} use the same categorization for filament morphology: L, C, S, X, and H. These classifications are defined as follows: L for linear filaments; C for bent, C-shaped filaments; S for quasi-sinusoidal filaments; X for bifurcated filaments; and H for hub-filament systems \citep{Wang2015}. Each filament is assigned no more than two morphological features based on the visually discerned shape of the MST spine or the $\rm N(H_2)$ maps. Figure \ref{fig1_F30}(a) illustrates the spine and surroundings of one selected Galactic LSF, F30, overlaid on the Hi-GAL $\rm N(H_2)$ map derived by \citet{Marsh2017}. 

\startlongtable
\begin{deluxetable*}{ccccccccc}
\tablenum{1}
\tablecaption{Properties of the 35 filaments\label{tab:35fils}}
\tablewidth{2pt}
\linespread{1.2}
\decimalcolnumbers
\tablehead{
\colhead{ID} & \colhead{$l_{wt}$} & \colhead{$b_{wt}$} 
& \colhead{$v_{wt}$} & \colhead{$d$} & \colhead{$\rm L_{sum}$} 
& \colhead{w} & \colhead{Morph} & \colhead{Reference}\\
\colhead{} & \colhead{(deg)} & \colhead{(deg)} 
& \colhead{($\rm km\ s^{-1}$)} & \colhead{(kpc)} & \colhead{(pc)} 
& \colhead{(pc)} & \colhead{} & \colhead{}}
\startdata
F1 & 5.37 & 0.13 & 11.0 & $\rm 2.9^{+0.3}_{-0.3}$ & 12.133 & 0.80 & L,C & I \\
F8 & 11.09 & -0.10 & 29.7 & $\rm 2.9^{+0.3}_{-0.3}$ & 22.809 & 0.88 & S & I \\
F19 & 15.65 & -0.22 & 57.0 & $\rm 11.6^{+0.3}_{-0.3}$ & 25.846 & 4.23 & C & I \\
F24 & 18.28 & -0.26 & 68.4 & $\rm 4.9^{+0.3}_{-0.3}$ & 15.547 & 1.24 & L & I \\
F30 & 22.40 & 0.35 & 84.0 & $\rm 5.4^{+0.3}_{-0.3}$ & 17.892 & 2.10 & L,X & I \\
F31 & 22.57 & -0.20 & 76.6 & $\rm 4.2^{+0.3}_{-0.3}$ & 12.65 & 1.86 & S & I \\
F34 & 22.75 & -0.46 & 76.4 & $\rm 4.6^{+0.3}_{-0.3}$ & 35.129 & 1.52 & S & I \\
F40 & 23.92 & 0.51 & 95.4 & $\rm 5.8^{+0.3}_{-0.3}$ & 53.725 & 3.26 & L & I \\
F45 & 25.36 & -0.38 & 57.1 & $\rm 2.7^{+0.3}_{-0.3}$ & 10.029 & 0.58 & L,X & I \\
F50 & 28.57 & -0.28 & 88.0 & $\rm 4.7^{+0.3}_{-0.3}$ & 15.900 & 1.50 & L,C & I \\
F56 & 32.03 & 0.07 & 95.6 & $\rm 5.2^{+0.3}_{-0.3}$ & 16.609 & 1.54 & S & I \\
F58 & 33.63 & -0.01 & 104.0 & $\rm 6.5^{+0.3}_{-0.3}$ & 50.951 & 2.18 & L & I \\
F62 & 41.16 & -0.21 & 60.0 & $\rm 8.9^{+0.3}_{-0.3}$ & 35.214 & 2.80 & S & I \\
F101 & 340.93 & -0.32 & -44.8 & $\rm 3.3^{+0.3}_{-0.3}$ & 15.445 & 1.42 & X & I \\
F109 & 337.70 & 0.10 & -75.2 & $\rm 10.8^{+0.3}_{-0.3}$ & 31.061 & 2.08 & C & I \\
F110 & 337.41 & -0.39 & -41.5 & $\rm 3.0^{+0.3}_{-0.3}$ & 10.729 & 0.84 & C & I \\
F113 & 337.77 & -0.34 & -41.2 & $\rm 3.0^{+0.3}_{-0.3}$ & 11.484 & 0.72 & X & I \\
F115 & 336.48 & -0.23 & -86.8 & $\rm 5.0^{+0.3}_{-0.3}$ & 33.020 & 1.54 & S & I \\
F126 & 332.36 & -0.06 & -48.4 & $\rm 3.1^{+0.3}_{-0.3}$ & 57.644 & 0.54 & C & I \\
F134 & 334.64 & 0.43 & -65.2 & $\rm 4.0^{+0.3}_{-0.3}$ & 14.858 & 2.11 & L,X & I \\
F135 & 345.50 & 0.34 & -17.1 & $\rm 2.4^{+0.3}_{-0.3}$ & 12.425 & 2.98 & S,X & I \\
F152 & 320.28 & -0.30 & -65.8 & $\rm 8.6^{+0.3}_{-0.3}$ & 41.655 & 0.97 & S & I \\
F153 & 320.40 & 0.13 & -4.9 & $\rm 12.5^{+0.3}_{-0.3}$ & 33.993 & 1.89 & C & I \\
F160 & 309.15 & -0.30 & -42.8 & $\rm 3.5^{+0.3}_{-0.3}$ & 16.159 & 1.34 & C,X & I \\
F4 & 8.16 & 0.22 & 19.8 & $\rm 2.9^{+0.28}_{-0.28}$ & 10.284 & 0.52 & L,H & II \\
F14 & 14.72 & -0.18 & 39.0 & $\rm 3.3^{+0.61}_{-0.61}$ & 14.925 & 1.76 & C,X & II \\
F23 & 23.37 & -0.12 & 97.5 & $\rm 5.9^{+0.13}_{-0.13}$ & 23.257 & 2.78 & S & II \\
F26 & 24.53 & -0.24 & 98.7 & $\rm 5.1^{+0.24}_{-0.24}$ & 18.834 & 1.36 & C & II \\
F28 & 25.30 & -0.22 & 63.3 & $\rm 3.9^{+0.85}_{-0.85}$ & 19.454 & 0.97 & L,H & II \\
G26L & 26.38 & 0.79 & 47.7 & $\rm 3.13^{+0.18}_{-0.19}$ & 31.337 & 0.92 & X & III \\
G26U & 26.38 & 0.79 & 47.7 & $\rm 3.13^{+0.18}_{-0.19}$ & 18.590 & 0.93 & X & III \\
G29 & 29.18 & -0.34 & 93.8 & $\rm 5.15^{+0.24}_{-0.23}$ & 22.893 & 2.00 & L & III \\
G47 & 47.06 & 0.26 & 57.5 & $\rm 4.44^{+0.56}_{-0.56}$ & 41.067 & 1.66 & C & III \\
G49 & 49.21 & -0.34 & 68.5 & $\rm 5.41^{+0.31}_{-0.28}$ & 33.037 & 1.77 & H,X & III \\
G64 & 64.27 & -0.42 & 22.0 & $\rm 3.62^{+1.56}_{-1.56}$ & 37.487 & 1.23 & L,H & III \\
G339 & 338.47 & -0.43 & -37.5 & $\rm 2.83^{+0.26}_{-0.28}$ & 60.323 & 0.80 & S,H & III \\
\enddata
\tablecomments{Column (1): ID of the LSFs. Most of them are directly adopted from references, except for G26 with a bifurcated structure, where ``L" stands for ``Lower Branch" and ``U" stands for ``Upper Branch". Columns (2)–(4): flux-weighted Galactic longitude, flux-weighted Galactic latitude (deg), and line-of-sight velocity ($\rm km\ s^{-1}$). Column (5): distance and distance uncertainty (kpc). Column (6): length of the MST filament spine (pc). Column (7): width of the filament (pc), defined as the FWHM of the $p=4$ Plummer-like fit to the segment-length-weighted average Hi-GAL $\rm N(H_2)$ radial profiles \citep{Suri2019}. Column (8): morphology class adapted from references (see Section \ref{sec:sourcedata}). Column (9): parent catalog of the filament. I for \citet{Ge2022}, II for \citet{Wang2016}, and III for \citet{Wang2015}.}
\end{deluxetable*}

\section{Method and Results} \label{sec:methodresults}
\subsection{Extraction of the Radial Profiles} \label{subsec:radprofmethod}
Extracting radial profiles requires sampling in the direction perpendicular to the filament spine. We first estimate the spine of each filament sample. For filaments identified using the MST algorithm, we take their major branch as the spine. For filaments from \citet{Wang2015}, we construct their spines by connecting local maxima in their Hi-GAL $\rm N(H_2)$ maps and ensure velocity coherence along these spines with $\rm ^{13}CO$ PV slices generated from SEDIGISM\citep{Schuller2021, Duarte-Cabral2021}, FUGIN\citep{Umemoto2017} and GRS\citep{Jackson2006}. Since the spines of our LSF samples can all be represented by a set of points and the segments connecting them, we treat each segment in the filament spine as a unit for radial profile extraction. Therefore, the radial profile sampling regions are rectangles on both sides of the filaments, perpendicular to the segments, with widths equal to the segment lengths. 

Next, we design the sampling length of the radial profiles. The central part of the radial profile corresponds to the densest region within the filament. It can not only take up a significant part in the calculation of asymmetry introduced in Section \ref{subsec:asy_param}, but also be easily affected by the choice of filament spine. To reduce the influence of spine accuracy on asymmetry calculations, we set the start of the radial profile sampling region as half of the filament width to avoid this central region and focus on the asymmetric feature in the surrounding environment of the filaments. 

Given that our 35 filaments reside at different distances with varied surroundings, assigning a universal sampling length across all LSF samples will increase the risk of including irrelevant surrounding structures. However, setting the sampling length as a constant $n$ times the filament width $w$, with a unified $n$ for all filaments also introduces problems. Filaments at far distances suffer from smoothing, a large $n$ would cause their radial profiles to include irrelevant structures, while a small $n$ would lead to insufficient sampling for nearby filaments. Therefore, we choose different $n$ values for different distances, using the filaments in the RCW120 system as references \citep{Zavagno2020}. These filaments have already been identified as hosting asymmetric density and temperature radial profiles generated by the central H\textsc{ii} region. The persistence of asymmetric features in their radial profiles can serves as a reference for determining the sampling length. 

We first apply Gaussian smoothing to the Hi-GAL $\rm N(H_2)$, $\rm T_d$, and their uncertainty maps of the RCW120 system to match all distances of our LSF samples, simulating the observed result of RCW120 filaments if they are located at those distances. The Hi-GAL maps are smoothed by convolving Gaussian kernels of appropriate sizes with the original images. Given the target distances of the LSFs and the distance to the RCW120 filaments ($\rm 1.34\ kpc$), the kernel size is computed as below:
\begin{equation}
{\rm FWHM_{kernel}}=\sqrt{{\frac{d^2-({1.34\rm\ kpc}^2)}{1.34\rm\ kpc^2}}}\times12\arcsec
\end{equation}
where $d$ denotes the distance of the LSFs, also the target distance to which the RCW120 image is smoothed. The 12\arcsec corresponds to the angular resolution of the Hi-GAL maps. 

\begin{figure*}[!t]
    \centering
    \includegraphics[width = 1.\textwidth]{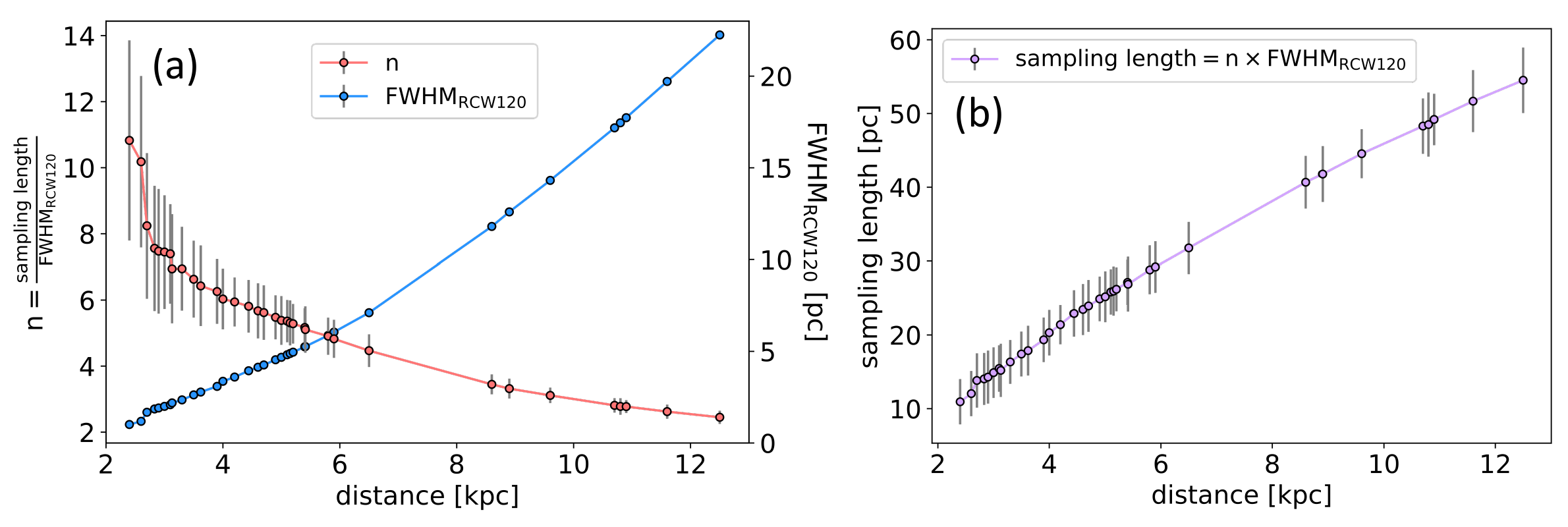}
    \caption{(a) Distribution of $n$, the ratio between the maximum asymmetry-recoverable sampling length and the Gaussian-smoothed FWHM of RCW120 filaments, and $\rm FWHM_{RCW120}$ at different distances. Each scattered point and error bar of $n$ corresponds to the mean and standard deviation among 1000 $n$ values, derived from mock radial profiles of RCW120 filaments at that distance. (b) Distribution of the sampling length, equal to $n\times \rm FWHM_{RCW120}$, versus LSF distance. The increasing trend of the curve highlights the importance of varying sampling lengths for LSFs at different distances. }
    \label{fig2_samplength}
\end{figure*}

Next, we define the observed width of the smoothed RCW120 filaments as the FWHM of the $p=4$ Plummer-like fit to their $\rm N(H_2)$ radial profiles extracted from the smoothed map. Then, we generate 1000 mock $\rm N(H_2)$ and $\rm T_d$ radial profiles of smoothed RCW120 filaments by sampling from a uniform distribution within their uncertainty ranges. The mock profiles simulate intrinsic asymmetric radial profiles with random perturbations, are used to constrain the upper limit of the sampling length, which is defined as the maximum radial extent at which an asymmetric profile can still be identified as asymmetric. For each mock profile, we set the start of the sampling region at half of the FWHM to avoid the central part, then gradually increase the upper limit of the sampling length one pixel at a time, until this mock profile can no longer be identified as asymmetric based on the criteria introduced in Section \ref{subsec:asy_param}. We record the ratio $n$ between this maximum asymmetry-recoverable sampling length to the width of smoothed RCW120 filament at that distance. The average $n$ from 1000 mock profiles is then used to define the sampling length of the LSF at that distance, starting at $0.5w$ and ending at $n\times w$. 
Figure \ref{fig2_samplength} shows the variation of the estimated ratio $n$, the FWHM of the smoothed RCW120 filaments, and the resulting sampling length as functions of LSF distance. Given the significant increase in FWHM with distance, the decreasing trend of $n$ can be interpreted as compensation for the smoothing of distant filaments. Moreover, the total sampling length $\rm n\times FWHM_{RCW120}$ is not constant across distances, this further emphasizes the importance of assigning different $n$ values when calculating sampling lengths. 

We draw each sampling region as a rectangular ``projection" region in DS9, and use the analyze function of the ``projection" region to extract the radial profiles of each segment in the LSF spine. The length and width parameters of each ``projection" region are set to the corresponding sampling length as $0.5w$ to $n\times w$ of the filament, and the segment length, respectively. The analyze function of the ``projection" region in DS9 averages the values of pixels at the same perpendicular distance from the segment, and repeats this process with an interval of one pixel until reaching the end of the sampling length, thereby generating the radial profile. By applying this method to each segment in the 35 LSFs on their $\rm N(H_2)$, $\rm T_d$ and uncertainty maps, we obtain $\rm N(H_2)$ and $\rm T_d$ radial profiles on both sides of each segment. Figure \ref{fig1_F30}(b) shows an example of the extracted, segment-length-weighted average $\rm N(H_2)$ and $\rm T_d$ radial profiles on both sides of LSF F30.

\subsection{Estimating the Asymmetric Features in Radial Profiles} \label{subsec:asy_param}

We introduce two parameters, $\alpha_{\rm asy}$ and $f_{\rm asy}$, to estimate the degree of asymmetry within the radial profiles of our LSFs, and the length proportion of segments with asymmetric radial profiles along the LSF spine, respectively.

The parameter $\alpha_{\rm asy}$ is defined as the ratio between the integrated areas beneath the radial profiles:
\begin{equation} 
\begin{split}
&\alpha_{\rm asy,N} =\frac{\sum^{n\times w}_{r=0.5w}\rm N_{Side1}(r)\Delta r}{\sum^{n\times w}_{r=0.5w}\rm N_{Side2}(r)\Delta r},\  \text{for $\rm N(H_2)$ case}\\
&\alpha_{\rm asy,T} =\frac{\sum^{n\times w}_{r=0.5w}\rm T_{Side1}(r)\Delta r}{\sum^{n\times w}_{r=0.5w}\rm T_{Side2}(r)\Delta r},\  \text{for $\rm T_d$ case}
\end{split}
\label{eq_alpha}
\end{equation}
where $\rm N(r)$ and $\rm T(r)$ represent the $\rm N(H_2)$ and $\rm T_d$ values at position $r$ in the radial profile. Side 1 generally refers to the northern/eastern side, and side 2 to the southern/western side. We calculate the value and uncertainty of $\alpha_{\rm asy, N}$ and $\alpha_{\rm asy, T}$ for each segment based on its radial profiles. If a segment has a completely symmetric environment, the radial profiles on both sides of the segment should be identical, resulting in $\alpha_{\rm asy} = 1$. We classify a segment as having a symmetric $\rm N(H_2)$ radial profile if its $\alpha_{\rm asy, N}$ overlaps with 1 within its uncertainty range (corresponding to $1 \sigma$), and similarly for the $\rm T_d$ radial profile.

Using $\alpha_{\rm asy}$, we identify segments with asymmetric radial profiles. We then calculate their length fraction $f_{\rm asy}$ within their parent filament spine:
\begin{equation} 
f_{\rm asy} = \frac{\sum_i{{\rm L}_{i,\ \alpha_{\rm asy} \neq 1}}}{{\rm L}_{\rm sum}}
\end{equation} 
where $\rm L_{sum}$ is the total length of the filament spine, and ${\rm L}_i$ is the length of segment $i$ whose $\alpha_{\rm asy}$ does not overlap with 1 within its uncertainty range, identified as having asymmetric surroundings. A large $f_{\rm asy}$ close to 1 indicates that almost all segments within the LSF spine have asymmetric surroundings. We calculate the $f_{\rm asy, N}$ and $f_{\rm asy, T}$ for each LSF based on the $\alpha_{\rm asy, N}$ and $\alpha_{\rm asy, T}$ values of each segment within it. 

\begin{figure}[!t]
    \centering
    \includegraphics[width = 0.48\textwidth]{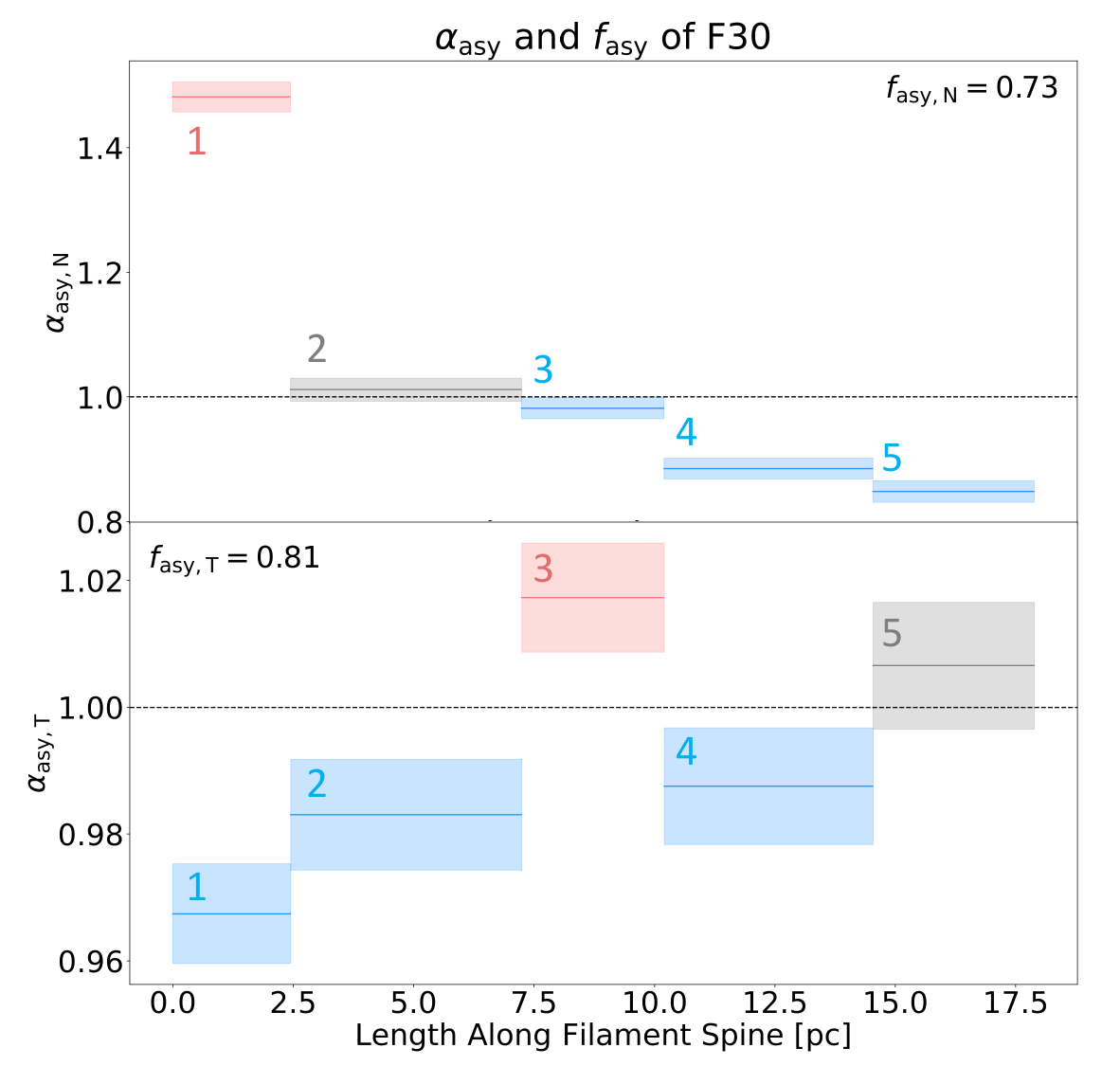}
    \caption{Distribution of the $\alpha_{\rm asy, N}$ and $\alpha_{\rm asy, T}$ for each segment in the spine of LSF F30 (same as Figure \ref{fig1_F30}), with the segment IDs labeled around them. The gray dashed line at a value of 1 represents the symmetric case. Segments highlighted in red or blue have $\alpha_{rm asy}$ that deviate from 1 within their uncertainty range and are thus identified as having asymmetric radial profiles. The values of $f_{\rm asy, N}$ and $f_{\rm asy, T}$ for this filament are displayed in the upper left or upper right corner of the panels. This filament has $f_{\rm asy, N} = 0.73$ and $f_{\rm asy, T} = 0.81$, meaning that 73\% in length of its spine has asymmetric $\rm N(H_2)$ radial profile, while 81\% in length of its spine has asymmetric $\rm T_d$ radial profile.}
    \label{fig3_asyF30}
\end{figure}

Figure \ref{fig3_asyF30} shows the $\alpha_{\rm asy, N}$ and $\alpha_{\rm asy, T}$ for each segment, as well as the $f_{\rm asy, N}$ and $f_{\rm asy, T}$ for LSF F30. The $\alpha_{\rm asy, N}$ values of segments 1, 3, 4, 5 do not overlap with 1 within their uncertainty range, so we classify them as having asymmetric $\rm N(H_2)$ radial profiles, resulting in the length proportion of asymmetric segments in the filament spine as $f_{\rm asy, N} = 0.73$. In the $\rm T_d$ case, the $\alpha_{\rm asy, T}$ values of segments 1-4 deviate from 1 within their uncertainty ranges, and we identify them as having asymmetric $\rm T_d$ radial profiles. Their length proportion in the entire LSF is 81\%, yielding $f_{\rm asy, T} = 0.81$.

\subsection{Unveiling Asymmetric Features in the Surroundings of Galactic LSFs} \label{subsec:asy_result}
\begin{figure*}[!t]
    \centering
    \includegraphics[width = 1.\textwidth]{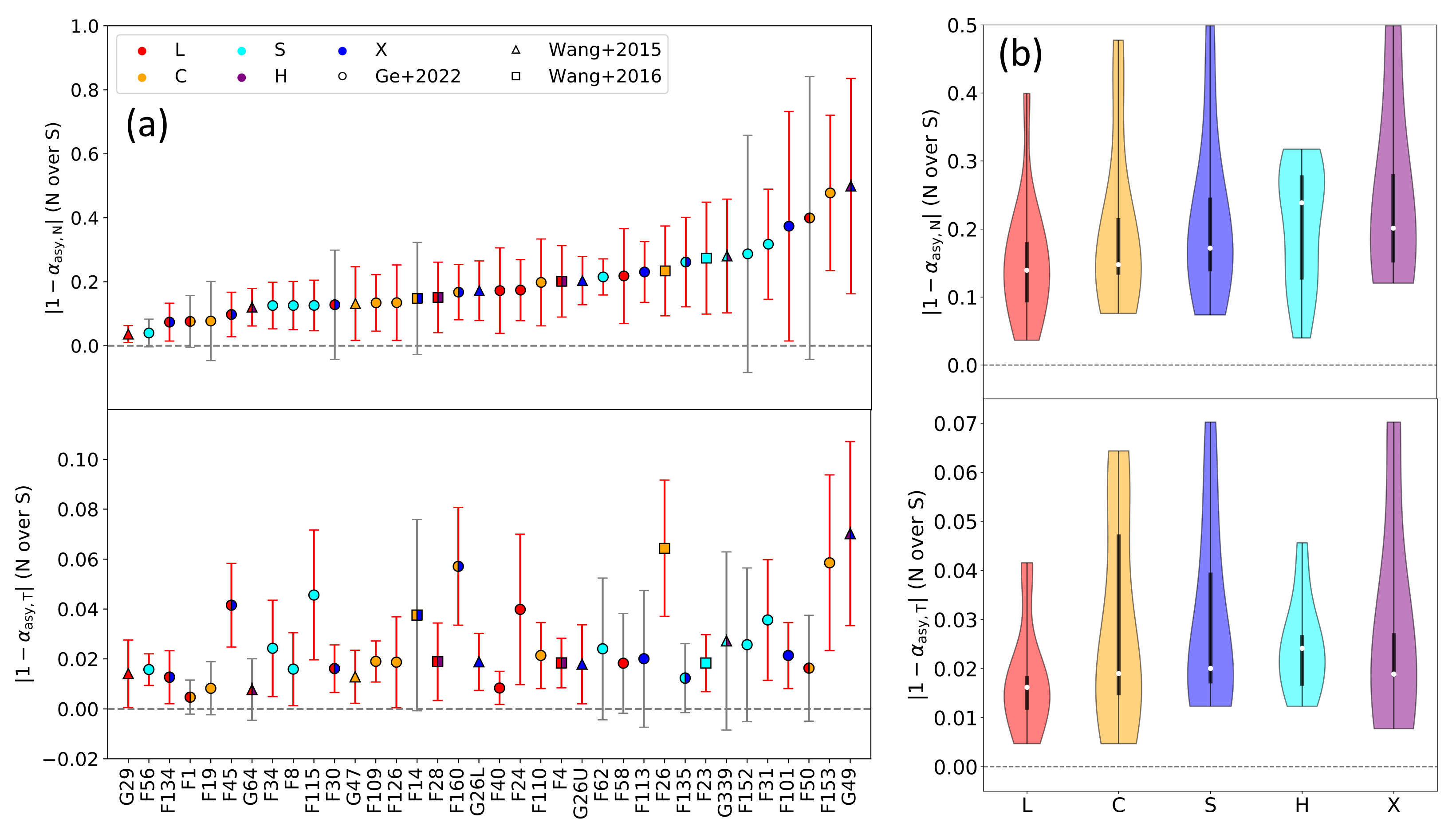}
    \caption{(a) Distribution of the weighted average asymmetric degree $|1-\alpha_{\rm asy}|$ in the $\rm N(H_2)$ and $\rm T_d$ radial profiles of the 35 filaments, weighted by the segment length. The upper panel corresponds to the $\rm N(H_2)$ case, the lower to the $\rm T_d$ case. The x-axis shows the filaments ordered by increasing $|1-\alpha_{\rm asy, N}|$. The shape of the data points indicates the parental catalog of the filament, and color represents the morphology of the filament: L (red), C (orange), S (cyan), H (purple), and X (blue), as defined in Section \ref{subsec:filsample}. Additional gray dashed horizontal line at 0 represents the symmetric scenarios. (b) Violin plots of $|1-\alpha_{\rm asy}|$ for the five LSF morphology classes. The upper panel shows the $\rm N(H_2)$ case, the lower panel shows the $\rm T_d$ case. The violins have their corresponding morphology types labeled in the x-axis, and their colors are in the same definition as (a). The white point in each violin represents the median, and the thick black line spans the 25th and 75th percentiles. Additional gray dashed horizontal line at 0 marks symmetric scenarios.}
    \label{fig4_alpha}
\end{figure*}

Figure \ref{fig4_alpha}(a) shows the weighted average $|1-\alpha_{\rm asy}|$ for each LSF, with the weights corresponding to the lengths of the segments within it. This value depicts the deviation of the LSF surroundings from the symmetric scenario. In $\rm N(H_2)$, 80\% (28 in 35) of the LSFs marked with red error bars show $|1-\alpha_{\rm asy, N}|$ deviating from 1, while in $\rm T_d$ the proportion is 68\% (24 in 35), indicating that Galactic LSFs generally reside in asymmetric surroundings of $\rm N(H_2)$ and $\rm T_d$. However, note that $|1-\alpha_{\rm asy, N}|$ has both values and total error bar lengths in the range of $\rm 0\sim1$, whereas for $|1-\alpha_{\rm asy, T}|$, the range is much smaller, only $0\sim 0.1$. The larger values and longer error bars of $|1-\alpha_{\rm asy, N}|$ indicate that while the $\rm N(H_2)$ distribution around the LSFs can sometimes differ markedly by several times of the lower side, it varies significantly among different positions surrounding the LSF. In contrast, the $\rm T_d$ distribution is relatively smooth and stable around the LSF, with differences commonly less than 10\% of the hotter side (i.e., about 2 K for a typical LSF temperature of 20 K). 

\begin{figure*}[!t]
    \centering
    \includegraphics[width = 1.\textwidth]{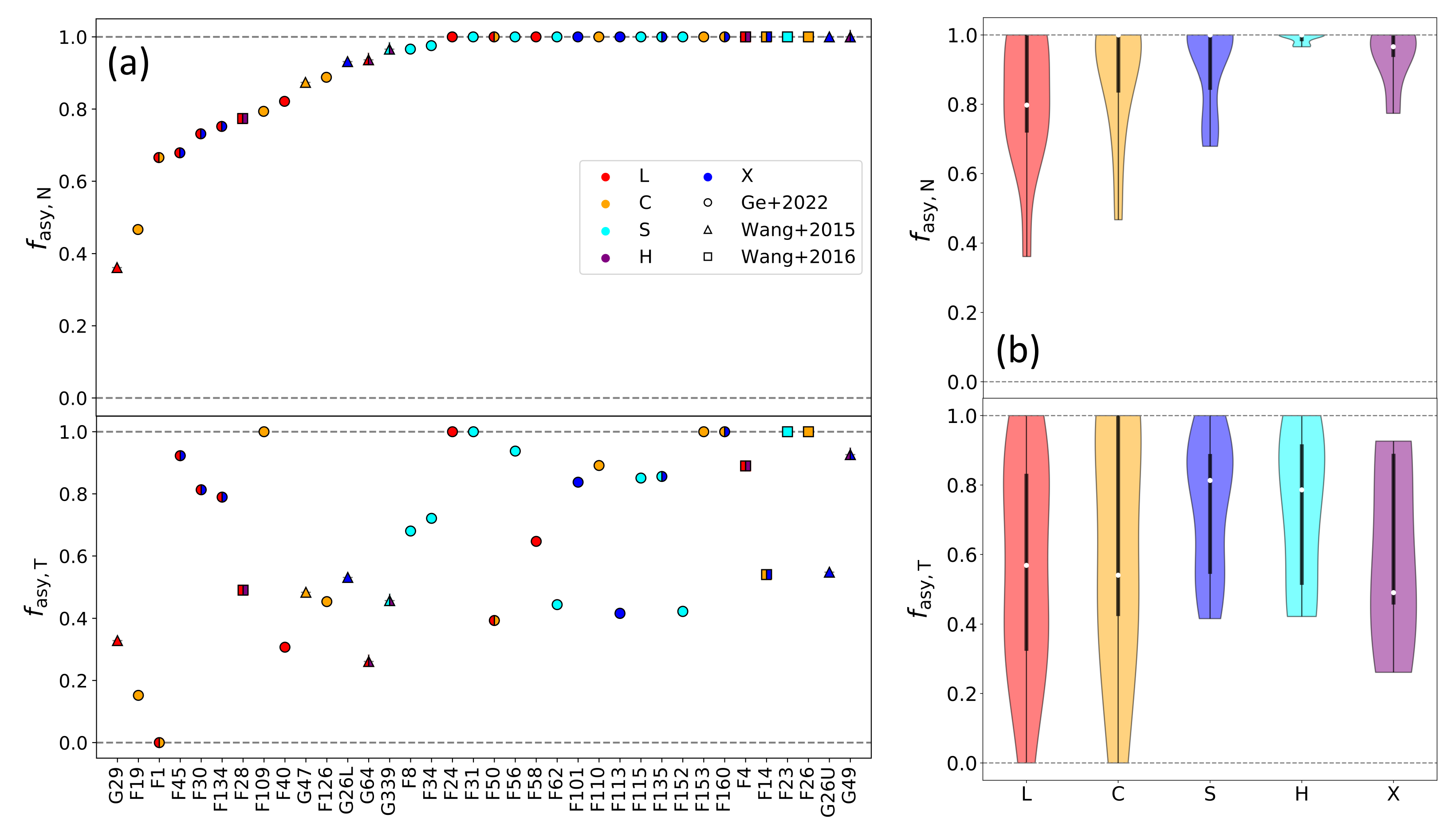}
    \caption{(a) Estimated $f_{\rm asy}$ based on the $\rm N(H_2)$ and $\rm T_d$ radial profiles of all 35 LSFs. The upper panel corresponds to the $\rm N(H_2)$ case, and the lower panel to the $\rm T_d$ case. The x-axis arranges the filaments by increasing $f_{\rm asy, N}$. Data point shapes and colors follow the same definitions as in Figure \ref{fig4_alpha}. Dashed gray dashed horizontal lines at 0 and 1 represent fully symmetric and fully asymmetric surroundings, respectively. (b) Violin plots of $f_{\rm asy}$ for the five LSF morphology classes, with the $\rm N(H_2)$ case in the upper panel, and the $\rm T_d$ case in the lower panel. Morphology labels and color conventions are the same as in Figure \ref{fig4_alpha}. Dashed gray horizontal lines at 0 and 1 represent fully symmetric and asymmetric surroundings, respectively.}
    \label{fig5_fasy}
\end{figure*}

On the other hand, Figure \ref{fig5_fasy}(a) shows the $f_{\rm asy}$ of each LSF. In the $\rm N(H_2)$ case, none of the LSFs have $f_{\rm asy,N} = 0$, and 20 LSFs have $f_{\rm asy,N} = 1$. In the $\rm T_d$ case, only one LSF has $f_{\rm asy,T} = 0$, while seven have $f_{\rm asy,T} = 1$. Almost all LSFs have more than half of their lengths residing in asymmetric $\rm N(H_2)$ environments, and a varied fraction of their lengths situated in asymmetric $\rm T_d$ surroundings. Together with the information obtained from $|1-\alpha_{\rm asy}|$, we conclude that Galactic LSFs generally reside in asymmetric $\rm N(H_2)$ and $\rm T_d$ surroundings, but with different behaviors. The $\rm N(H_2)$ distribution shows pronounced contrasts across the two sides of almost the entire LSF, and varies significantly for different parts of the LSF. In contrast, the $\rm T_d$ distribution exhibits milder differences across the sides of the LSF, and have relatively little variation along its length. 

The violin plots in Figure \ref{fig4_alpha}(b) and Figure \ref{fig5_fasy}(b) categorize the distribution of $|1-\alpha_{\rm asy}|$ and $f_{\rm asy}$ by different LSF morphology types. Recalling the H\textsc{ii} region-triggered formation of bent filaments, a natural hypothesis is that bent LSFs with ``C-shape" or ``S-shape" morphology may have experienced this mechanism, while straighter, more symmetric ``L-shape" LSFs are less related to it. If this assumption holds, we would expect C-shape and S-shape LSFs to have relatively more asymmetric surroundings, and L-shape LSFs to have more symmetric ones. However, in both the $\rm N(H_2)$ and $\rm T_d$ scenarios, the five LSF morphologies all have overlapping interquartile ranges of $|1-\alpha_{\rm asy}|$ and $f_{\rm asy}$. Although L-shape LSFs indeed have the smallest medians for the two cases of $|1-\alpha_{\rm asy}|$ and $f_{\rm asy,N}$, the C-shape and S-shape filaments are not those with the largest values. LSFs of different shapes do not differ markedly in either the degree of surrounding asymmetry or the length fraction of asymmetric segments. While a linear LSF shape may correlate with more symmetric surroundings and a lower fraction of spine with asymmetric radial profiles, a bent shape does not necessarily correspond to more asymmetric surroundings or a greater fraction of asymmetric segments. This contradicts the causal link suggested in the H\textsc{ii} region-triggered filament formation scenario.

Furthermore, the increasing trend of $|1-\alpha_{\rm asy, N}|$ seen in the upper panel of Figure \ref{fig4_alpha} (a) is not preserved in the lower panel showing $|1-\alpha_{\rm asy, T}|$. Similarly, the gradually increasing trend of $f_{\rm asy, N}$ in the upper panel of Figure \ref{fig5_fasy} (a) is not maintained in the lower panel showing $f_{\rm asy, T}$. Moreover, only 6 LSFs have $f_{\rm asy,N} = f_{\rm asy,T}$. These results demonstrate that segments with surrounding identified as asymmetric in $\rm N(H_2)$ are not necessarily identified as asymmetric in $\rm T_d$. Given this discrepancy between density and temperature asymmetry identification, it becomes unlikely that a single mechanism simultaneously producing both types of asymmetry is the dominant driver of Galactic LSF formation. Combined with the weak correlation between bent LSF morphology and radial profile asymmetry, the H\textsc{ii} region-triggered filament formation scenario seems unlikely to be the major formation mechanism behind Galactic LSFs (further discussion on the spatial relationship between H\textsc{ii} regions and filaments is made in Section \ref{sec:discussion}). 

\begin{figure*}[!t]
    \centering
    \includegraphics[width = 1.\textwidth]{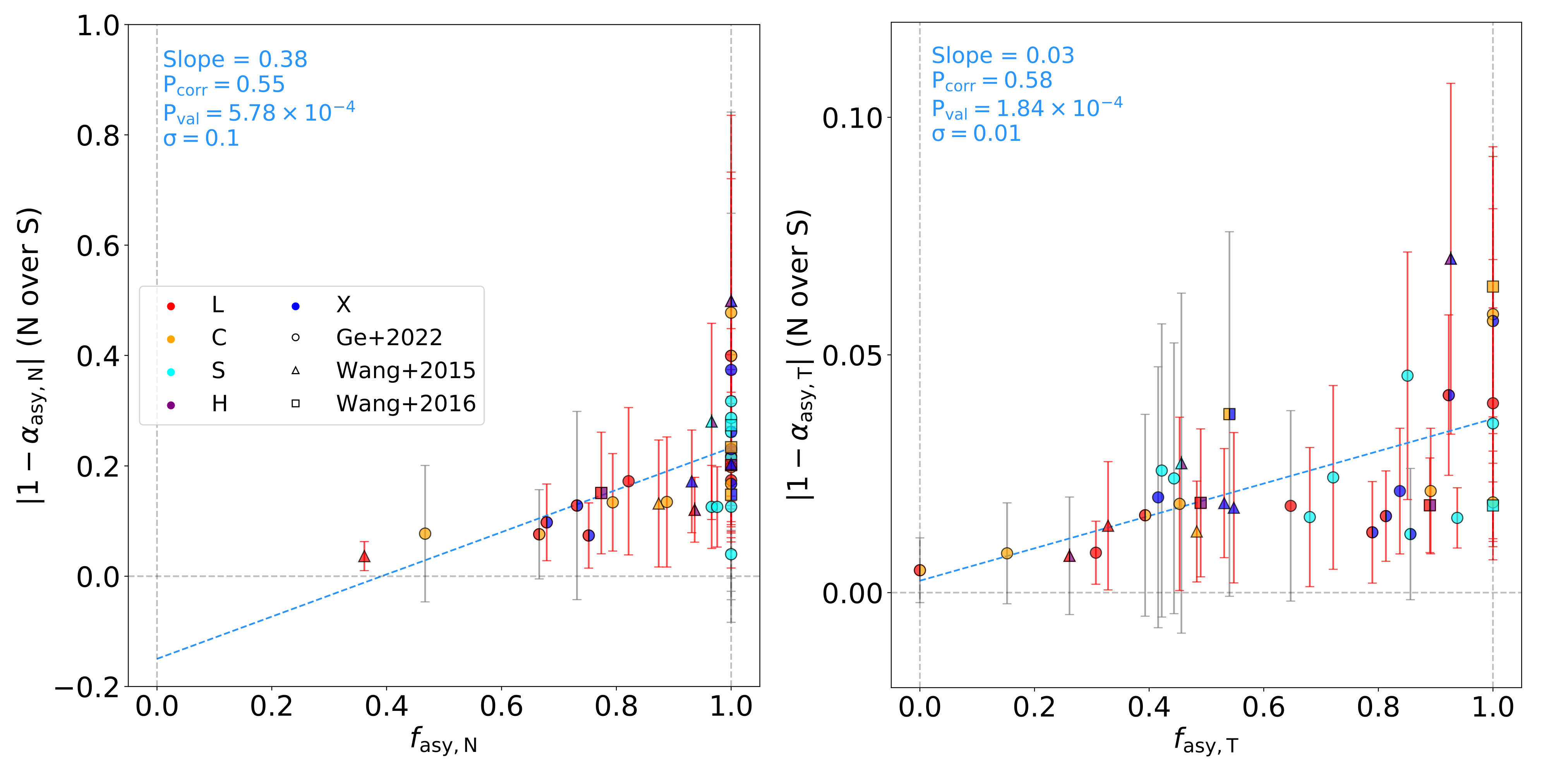}
    \caption{Distribution of segment-length-weighted average $|1-\alpha_{\rm asy}|$ versus $f_{\rm asy}$ of all 35 LSFs. The left panel shows the $\rm N(H_2)$ scenario, the right shows the $\rm T_d$ scenario. Data point shapes and colors follow the same definitions as in Figure \ref{fig4_alpha}. Gray dashed vertical lines at 0 and 1 indicate filaments with entirely symmetric or asymmetric surroundings, respectively. A gray dashed horizontal line at 0 marks filaments with fully symmetric weighted-average radial profiles. }
    \label{fig6_alphavsf}
\end{figure*}

Figure \ref{fig6_alphavsf} shows the distribution of segment-length-weighted average $|1-\alpha_{\rm asy}|$ for each LSF as a function of $f_{\rm asy}$. Proportional relationships are found between $|1-\alpha_{\rm asy}|$ and $f_{\rm asy}$ in both the $\rm N(H_2)$ and $\rm T_d$ scenarios. LSFs with error bars labeled in red, identified as having asymmetric surroundings tend to cluster at large $f_{\rm asy}$ values. More importantly, there are no LSFs with $f_{\rm asy} < 0.2$ (i.e., less than 2 pc for LSFs of lengths 10 pc) that are identified as having asymmetric surroundings. Together, these results indicate that the degree of asymmetry in LSF surroundings and the length proportion of asymmetric segments in LSF spines complement each other. None of our LSFs are found to have short segments with extreme $|1-\alpha_{\rm asy}|$ values large enough to alter the asymmetry classification result of the entire filament. Additionally, the slope of the proportional relationship in the $\rm N(H_2)$ case is steeper than that in the $\rm T_d$ case. Almost all LSFs with asymmetric $|1-\alpha_{\rm asy,N}|$ cluster at large $f_{\rm asy,N}$, while LSFs with asymmetric $|1-\alpha_{\rm asy,T}|$ distribute relatively more uniform across different $f_{\rm asy}$. These trends are consistent with the distinctly asymmetric and variable $\rm N(H_2)$ environments, and the milder, more stable asymmetric $\rm T_d$ surroundings of Galactic LSFs.

\section{Discussion} \label{sec:discussion}
\begin{figure*}[!t]
    \centering
    \includegraphics[width = 1.\textwidth]{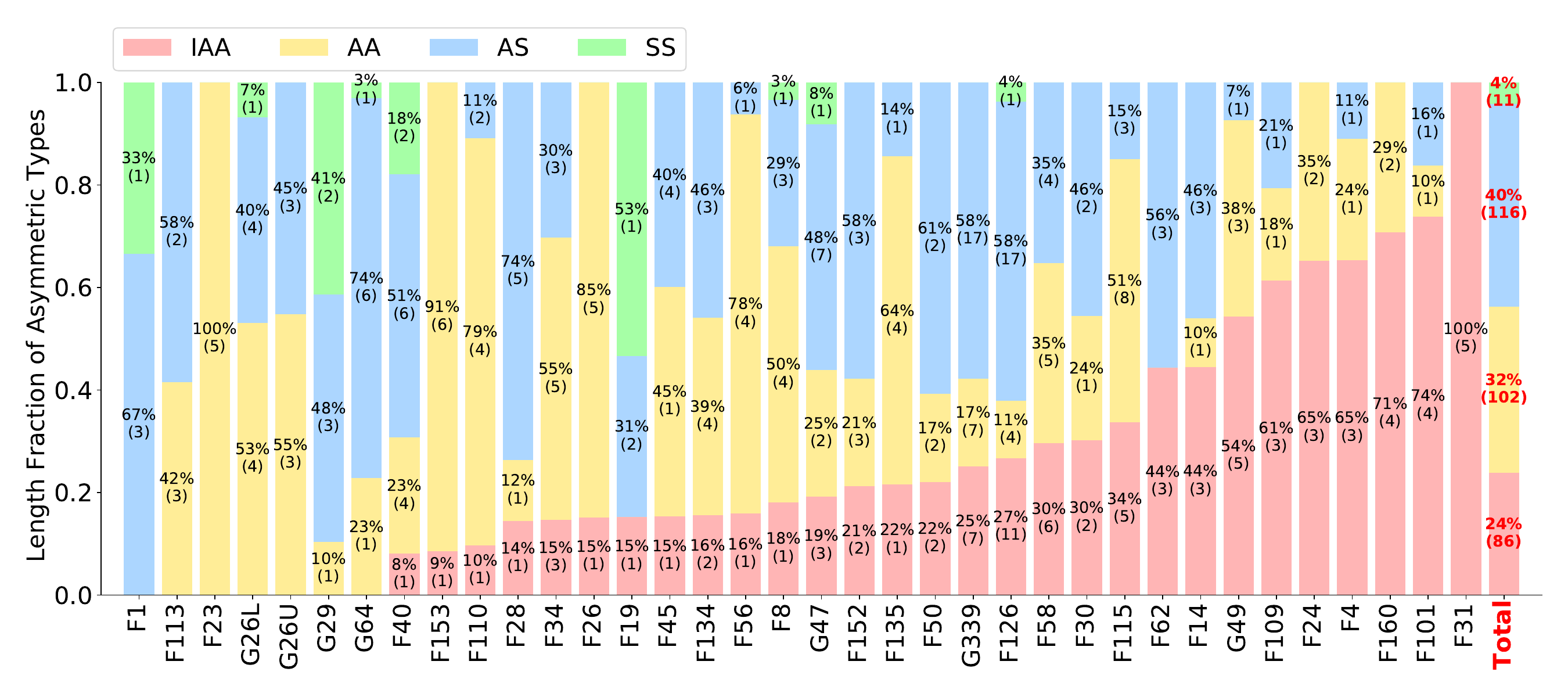}
    \caption{Asymmetric categorization result of the segments, and the length proportions of each asymmetric types in the 35 LSFs and in the total length of the 35 LSFs. Colors of red, yellow, blue, and green represent asymmetric types IAA, AA, AS, and SS, respectively. The percentage values and the numbers in brackets inside each bar denote the length proportion of segments with the corresponding asymmetric type, and the number of segments of that type in the LSF. The x-axis lists the filaments in order of increasing length proportion of type IAA segments.}
    \label{fig7_filasytype}
\end{figure*}

Differences between the trends in Figure \ref{fig4_alpha}(a) and Figure \ref{fig5_fasy}(a) imply the possibility that some segments have surroundings identified as asymmetric in only one type of $\rm N(H_2)$ or $\rm T_d$, which deviates from the H\textsc{ii} region-triggered filament formation scenario described in Section \ref{sec:intro} \citep{Inutsuka2015, Pineda2023}. To assess the frequency of this deviation, we first examine the asymmetric identification results of all 315 segments within the 35 LSFs. We categorize the segments into 4 types: (1) Asymmetric in both $\rm N(H_2)$ and $\rm T_d$ radial profiles, with an inverse trend (IAA), meaning the denser side is cooler. This category aligns with the H\textsc{ii} region-triggered filament formation mechanism. (2) Asymmetric in both $\rm N(H_2)$ and $\rm T_d$ radial profiles, but with the same trend (AA), meaning the denser side is hotter. (3) Asymmetric in only one of $\rm N(H_2)$ or $\rm T_d$ (AS). (4) Symmetric in both cases (SS). The categorization results are shown in Figure \ref{fig7_filasytype}. Most of the segments fall into the AA and AS type, their lengths account for 72\% of the total length of the 35 LSFs. Segments of type IAA take varied length proportions in 28 out of the 35 LSFs, constituting about one quarter of the total LSF lengths. The SS type is the least common, comprising only 11 segments and occupying 4\% of the total LSF lengths.

Next, we investigate whether nearby H\textsc{ii} regions contribute to the asymmetric features of the IAA segments by analyzing their spatial relationship. We identify H\textsc{ii} regions using a combination of the MeerKAT spectral index maps \citep{Goedhart2024} and the WISE-based Galactic H\textsc{ii} region catalog \citep{Anderson2014}. A structure is identified as an H\textsc{ii} region if it not only has a spectral index slightly less than 0 in the MeerKAT spectral index map \citep{Makai2017}, but is also covered by at least one identified H\textsc{ii} region from the WISE catalog. An IAA segment is considered to possibly influenced by a nearby H\textsc{ii} regions if it meets both of the following criteria: (1) its radial profile sampling regions overlap with at least one H\textsc{ii} region, and preferably, the segment is located at the edge of that H\textsc{ii} region; (2) the line-of-sight distance of its parental LSF matches the H\textsc{ii} region in the uncertainty range.

We plot the LSFs and nearby WISE H\textsc{ii} regions on the MeerKAT spectral index map, highlighting the IAA segments in magenta and labeling the radial profile sampling regions of each segment. A full set of plots for all 35 LSFs is shown in Appendix \ref{App_fil+si+HII}. Figure \ref{fig8_IAAfil} displays the three LSFs having IAA segments that meets all of the above criteria, along with the Hi-GAL $\rm N(H_2)$ and $\rm T_d$ distribution around the LSF. All of the IAA segments in F62, the middle IAA segment of G47, and the IAA segment at the western end of G49 are identified as associated with H\textsc{ii} regions, with nearby H\textsc{ii} regions at similar distances. The H\textsc{ii} regions all appear as dense, arc-like $\rm N(H_2)$ rim, circular $\rm T_d$ enhancement, and circular region with a spectral index close to 0. The associated IAA segments reside on the edge of the H\textsc{ii} regions, overlapping with the arc-like $\rm N(H_2)$ structure. Although kinematic data is needed to confirm whether the segments are indeed experiencing feedback from the H\textsc{ii} region, the above evidence, along with the matched distances, strongly suggests that the surrounding density and temperature distribution of the segments were likely shaped by the H\textsc{ii} region.

Aside from the three LSFs containing IAA segments identified as associated with H\textsc{ii} regions, we also found three LSFs with IAA segments that could possibly be influenced by nearby H\textsc{ii} regions. We plot their spines, nearby H\textsc{ii} region distributions, and sampling regions on their corresponding MeerKAT spectral index, Hi-GAL $\rm N(H_2)$ and $\rm T_d$ maps in Figure \ref{fig9_likelyIAAfil}. The most west IAA segment of F24, the IAA segment in the middle of F152, and the IAA segments of G339 , could possibly be influenced by H\textsc{ii} regions marked by red arrows. The orientation of the segments follows the rim of the H\textsc{ii} regions, and the $\rm N(H_2)$ and $\rm T_d$ distributions around them show theoretical arc-like shapes and circular enhancements. However, due to unknown distances or distance uncertainties, we cannot confirm their association, and thus label these segments as candidates possibly influenced by H\textsc{ii} regions. For other IAA segments in the remaining 22 LSFs, either their sampling regions lack circular structures with spectral indices near 0, or their asymmetric features result from complex structures within their parental molecular cloud. In such cases, the absence of H\textsc{ii} regions with shape comparable to their asymmetric surroundings suggests that their asymmetric features are unlikely to be caused by H\textsc{ii} region feedback.

The presence of IAA segments in 28 out of the 35 LSFs, accounting for one quarter of the total segment length, indicates that it is common for parts of LSFs to reside in environments where both $\rm N(H_2)$ and $\rm T_d$ are asymmetric and have opposite trends. However, among the 28 LSFs hosting IAA segments, only three have IAA segments identified as being influenced by nearby H\textsc{ii} regions, accounting for 44\% (F62), 4\% (G47), and 7\% (G49) of their respective total lengths. The majority of IAA segments are instead attributed to the complex internal structure of the molecular clouds themselves. These findings suggest that H\textsc{ii} regions with sizes comparable to LSFs can indeed influence Galactic LSFs, as illustrated in Figure \ref{fig8_IAAfil}, but they are not the sole cause of the IAA environments observed around LSF segments. At the same time, the extent over which an H\textsc{ii} region can exert influence within an LSF is limited. Therefore, we conclude that H\textsc{ii} regions are unlikely to trigger the formation of entire Galactic-scale filaments, in contrast to the smaller, parsec-scale cases described by \citet{Zavagno2020}.

\begin{figure*}[!t]
    \centering
    \includegraphics[width = 1.\textwidth]{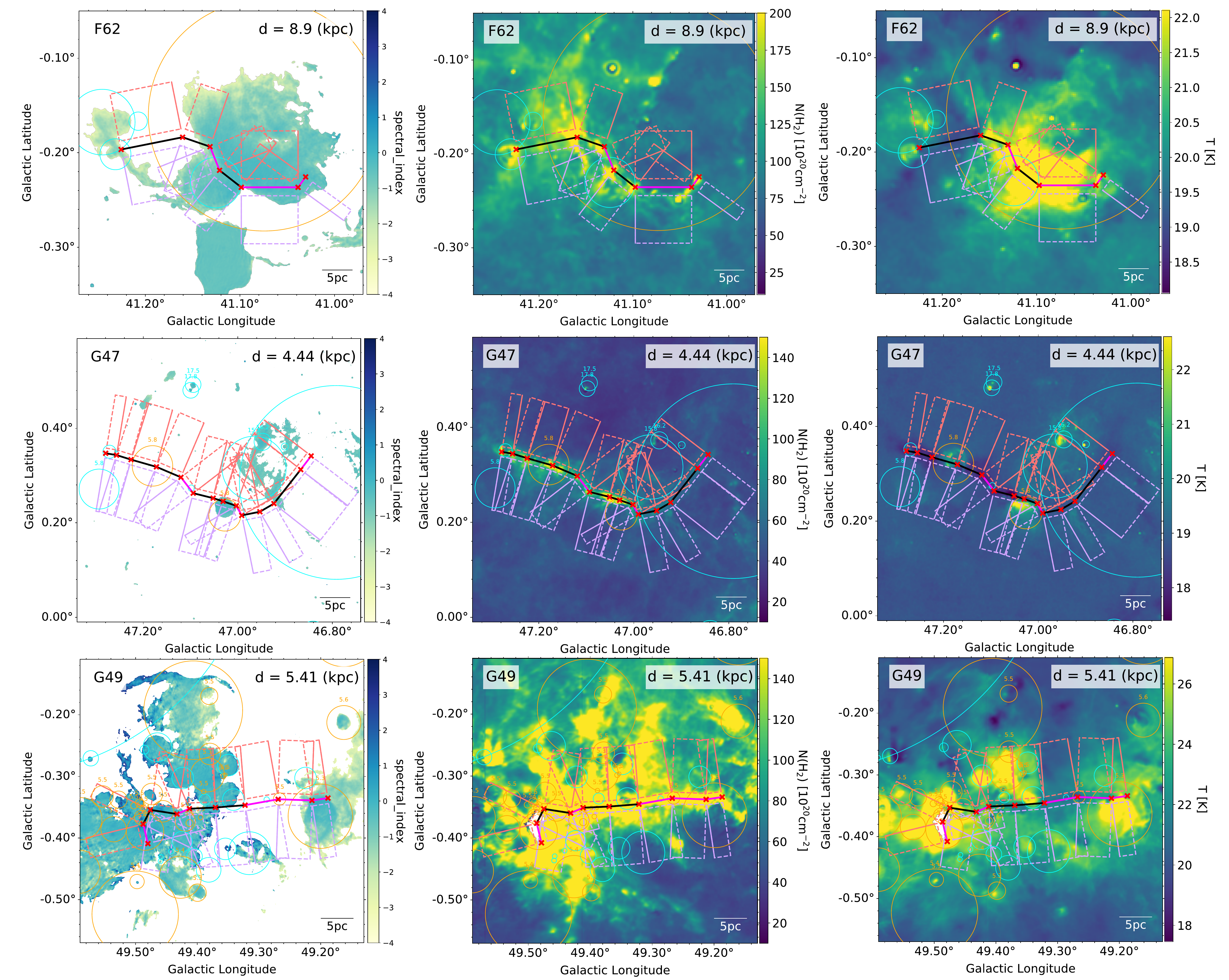}
    \caption{The three LSFs with IAA segments identified as influenced by nearby H\textsc{ii} regions. Backgrounds from the left to right column refer to the MeerKAT spectral index map \citet{Goedhart2024}, Hi-GAL $\rm N(H_2)$ and $\rm T_d$ map \citep{Marsh2017}. The red crosses and black segments connecting them represent the filament spine, segments colored in magenta have IAA radial profiles. Dashed boxes mark the radial profile sampling regions of each segment. Circles represent nearby WISE H\textsc{ii} regions, orange ones matches the filament distance within its uncertainty range, while cyan circles either differ in distance or lack distance information. A 5 pc scale bar is included in the lower right corner of each panel.}
    \label{fig8_IAAfil}
\end{figure*}

\begin{figure*}[!t]
    \centering
    \includegraphics[width = 1.\textwidth]{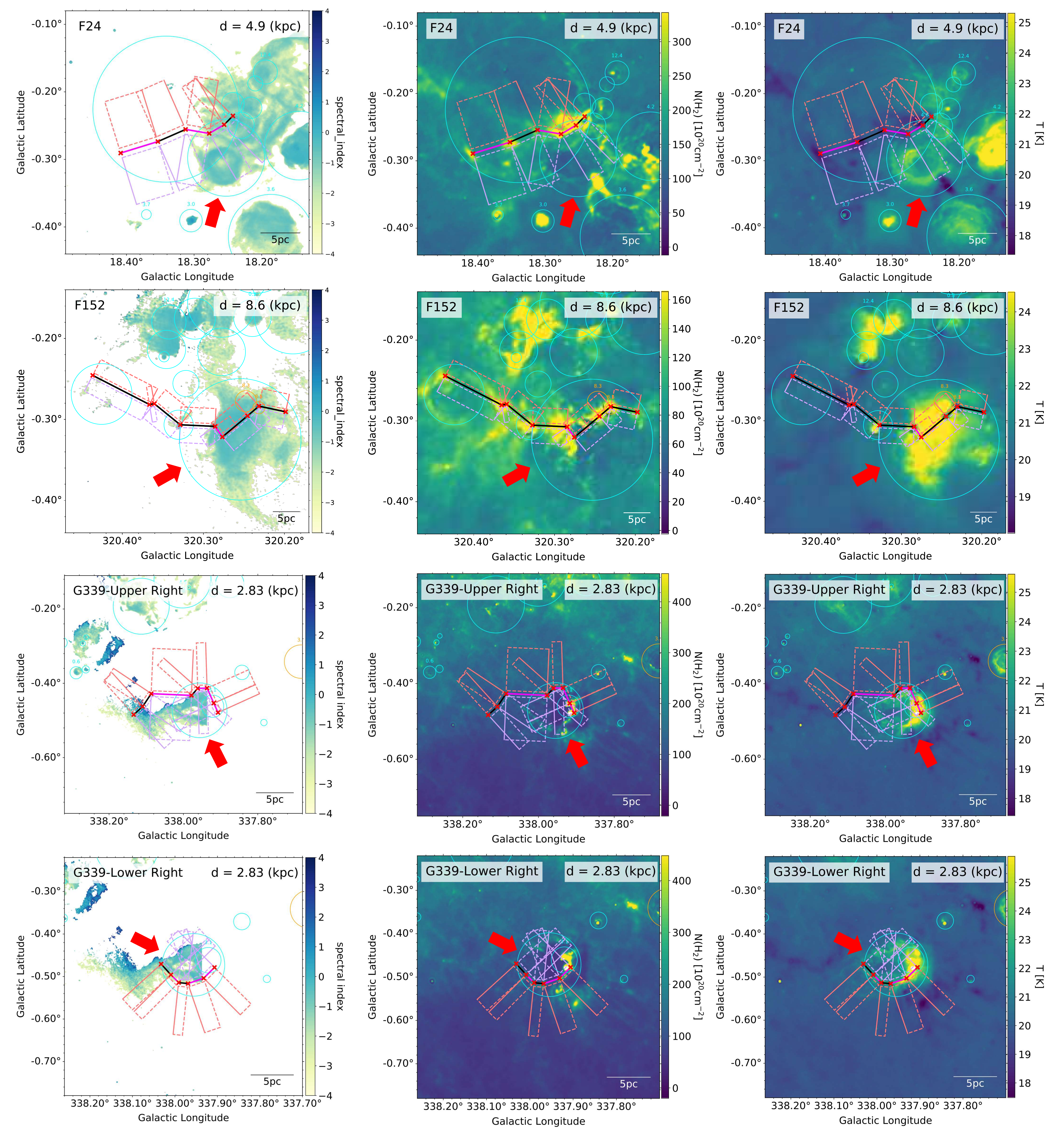}
    \caption{The three LSFs with IAA segments that possibly be influenced by nearby H\textsc{ii} regions. The background, crosses, lines, boxes, and circles have the same meanings as in Figure \ref{fig8_IAAfil}. Red arrows highlight the H\textsc{ii} regions that are likely to affect the IAA segments. Their potential effects remain unconfirmed due to unknown or uncertain distance measurements. For a clearer view of the sampling regions and surroundings of the IAA segments in G339, we present the relevant IAA segments in G339 in two sets of figures, their corresponding fields of view are shown in Figure \ref{fig_app1_spindex}. }
    \label{fig9_likelyIAAfil}
\end{figure*}

Unlike previous studies that analyze the radial profile averaged over the entire filament \citep{Zavagno2020, Clarke2023}, we divided $>$ 10 pc Galactic LSFs into smaller pc-scale segments and analyzed their radial profiles separately. Since H\textsc{ii} regions are generally smaller or comparable in size to Galactic LSFs, our approach avoids diluting the asymmetric features through averaging, and allows us to identify cases where only part of the filament is influenced. Furthermore, this segmentation method enables quantification of the length proportion of segments influenced by H\textsc{ii} regions, offering a complementary way to estimate the impact of H\textsc{ii} regions on Galactic LSFs beyond the averaged radial profile of the entire LSF. However, it is worth noting that the radial profiles we extracted through DS9 are still averaged over the pc-scale segments. This may result in the dilution of asymmetric features in LSF surroundings with sizes smaller than the pc scale.

It is worth noticing that LSFs can pass through H\textsc{ii} regions of varying sizes. As seen in Figure \ref{fig8_IAAfil} and Figure \ref{fig9_likelyIAAfil}, the radii of H\textsc{ii} regions range from a few pc to several tens of pc. Using asymmetric radial profiles to identify H\textsc{ii} region-triggered filament formation heavily depends on the chosen sampling length. Factors such as the irregular shape of the molecular cloud, other mechanisms that can cause interstellar medium to aggregate (e.g., cloud cloud collision \citet{Inutsuka2015, Pineda2023}) and limited resolution may prevent precise alignment of the filament spine with the H\textsc{ii} region rim. If the sampling length is just sufficient to place one side of the LSF overlapping with the H\textsc{ii} region and the other outside, as in the IAA segments of Figure \ref{fig8_IAAfil}, the asymmetry can be clearly detected. However, if the sampling length does not match the H\textsc{ii} region's scale, asymmetries may be missed—even if gas kinematics confirm H\textsc{ii} region feedback. In this study, the radial profile sampling length is designed referring to the RCW120 system, where the H\textsc{ii} region is similar in size to the filaments. Thus, our sampling length is only appropriate for examining the relationship between LSFs and H\textsc{ii} regions with sizes comparable to the LSF. For H\textsc{ii} regions much smaller or larger than the LSFs, different sampling scales would be required to adequately capture the induced asymmetries and reveal the filament–H\textsc{ii} region relationship.

\section{Conclusion} \label{sec:conclusion}
We selected 35 Galactic LSFs (length $\rm \geq 10\ pc$) from the filament catalogs of \citet{Ge2022}, \citet{Wang2016}, and \citet{Wang2015}. Their spines were divided into 315 segments, and the $\rm N(H_2)$ and $\rm T_d$ radial profiles were extracted for each segment. The $\rm N(H_2)$ and $\rm T_d$ maps used for extracting radial profiles were constructed based on $Herschel$ 70, 160, 250, 350, and 500 $\mu \rm m$ continuum images \citep{Marsh2017}. We defined two parameters, $\alpha_{\rm asy}$ and $f_{\rm asy}$, to identify the segments with asymmetric radial profiles and to estimate the length proportion of asymmetric surroundings in the LSFs, respectively. Using these parameters, we identified segments with surroundings asymmetric in $\rm N(H_2)$ and $\rm T_d$ distributions, calculated their length proportions to the filament spines, and compared their distribution with nearby H\textsc{ii} regions, traced by the spectral index maps of \citet{Goedhart2024} and the WISE H\textsc{ii} region catalog \citep{Anderson2014}. The main results are as follows:

1. Galactic LSFs reside in asymmetric surroundings of $\rm N(H_2)$ and $\rm T_d$. The $\rm N(H_2)$ distribution often exhibits pronounced asymmetry and varies significantly along the LSF, whereas the $\rm T_d$ distribution is comparatively milder and smoother. The segment-length-weighted average $|1-\alpha_{\rm asy}|$ quantifies the deviation of the radial profile from symmetry. All 35 LSFs show asymmetric $|1-\alpha_{\rm asy}|$ values in both cases, with the $\rm N(H_2)$ case displaying larger values and longer error bars than the $\rm T_d$ case. Nearly all LSFs have more than half of their total length embedded in asymmetric $\rm N(H_2)$ surroundings, and varying fractions of their spines in asymmetric $\rm T_d$ environments. These results reflect the distinctly asymmetric and variable $\rm N(H_2)$ distributions, and the relatively mild and stable asymmetric $\rm T_d$ distributions, characterizing the Galactic LSF surroundings.

2. Different LSF morphology types exhibit similar degrees of asymmetry in their surroundings, as well as similar length proportions of asymmetric segments within their spines. Contrary to the H\textsc{ii} region-triggered filament formation scenario, which predicts that bent LSFs are likely generated by this mechanism, should have more asymmetric radial profiles, our LSFs with bent ``C-shape" or ``S-shape" are neither having the largest $|1-\alpha_{\rm asy}|$ nor the largest $f_{\rm asy}$. Nevertheless, the straight ``L-shape" LSFs, predicted to be the least related to the H\textsc{ii} region-triggered filament formation scenario, indeed have small $|1-\alpha_{\rm asy}|$ and $f_{\rm asy}$. A bent LSF shape does not necessarily corresponds to a more asymmetric radial profile, but a linear, more symmetric morphology might indicate more symmetric surroundings. 

3. Segments with radial profiles identified as asymmetric in $\rm N(H_2)$ are not necessarily asymmetric in $\rm T_d$, and vice versa. Segments along the LSF spines commonly have surroundings that are asymmetric in both $\rm N(H_2)$ and $\rm T_d$ with the same trend, or are asymmetric in only one of them. 

4. Segments with IAA radial profiles, consistent with the H\textsc{ii} region-triggered filament formation scenario, account for about one quarter of the total length of the 35 LSFs. However, most IAA segments likely arise from the complex structure of the molecular clouds rather than from the influence of H\textsc{ii} regions. Only three LSFs have an averaged value of 18\% of their spines identified as containing IAA features that are likely generated by nearby H\textsc{ii} regions.  H\textsc{ii} regions with sizes comparable to Galactic LSFs can indeed induce inverse asymmetric radial profiles in Galactic LSFs. However, unlike the case for small-scale filaments, they are unlikely to trigger the formation of an entire Galactic LSF. 

\vspace{5mm}
\software{MST Filaments \citep{Wang2021_MST_code},
Astropy \citep{astropy:2013, astropy:2018, astropy:2022}, DS9 \citep{SAO2000, Joye2003}, pyds9}

\begin{acknowledgements}
We thank the anonymous referee for the thorough review and critical comments that helped us to improve the scientific content and the presentation of this paper. 
We acknowledge support from National SKA Program of China (2025SKA0140100), National Natural Science Foundation of China (No. 12573025), China-Chile Joint Research Fund (CCJRF No. 2211), the Tianchi Talent Program of Xinjiang Uygur Autonomous Region and the High-Performance Computing Platform of Peking University.
CCJRF is provided by Chinese Academy of Sciences South America Center for Astronomy (CASSACA) and established by National Astronomical Observatories, Chinese Academy of Sciences (NAOC) and Chilean Astronomy Society (SOCHIAS) to support China-Chile collaborations in astronomy.
\end{acknowledgements}

\appendix
\section{Spectral index distribution in the surroundings of the 34 LSFs}\label{App_fil+si+HII}

We have already shown the distribution of filaments with IAA segments in Figure \ref{fig8_IAAfil} and Figure \ref{fig9_likelyIAAfil}. Here, we present the spectral index surroundings of 34 LSFs (the only exception is G64, it was not covered in the region of MeerKAT), with their segments identified as having IAA features highlighted in magenta, along with nearby H\textsc{ii} regions identified in the WISE catalog \citep{Anderson2014}.

\renewcommand\thefigure{\Alph{section}\arabic{figure}} 
\setcounter{figure}{0} 
\begin{center}
\setlength{\tabcolsep}{1.2mm}{
{
    \begin{figure}
    \includegraphics[width=1.\linewidth]{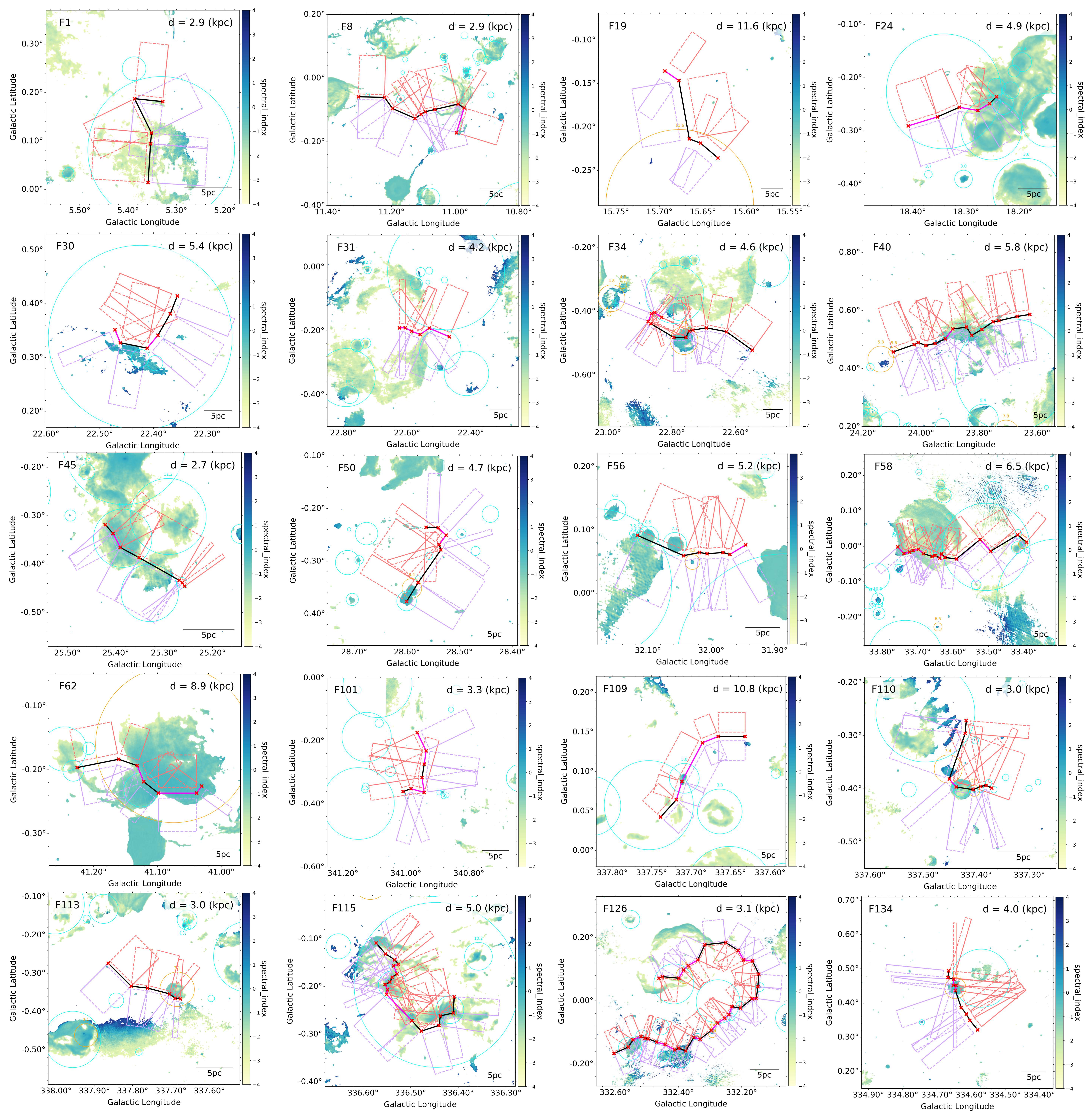}
    \caption{The 34 LSFs plotted on the spectral index map generated by \citet{Goedhart2024}. G64 is not included as it does not overlap with the field of view of the SMGPS. G339 is divided into 4 images for a clearer view of the sampling regions, shown in the last row. The field of view of each image is indicated by a black box with corresponding notes in the full image of G339. The crosses and line segments represent the spines of the filaments, with the IAA segments highlighted in magenta. The yellow dashed boxes outline the sampling regions of the radial profiles for each segment. The circles represent nearby H\textsc{ii} regions selected from the WISE catalog. Orange ones overlap with the distance of the filament within its uncertainty range, while cyan circles either differ in distance or lack distance information. A scale bar of 5 pc is labeled in the lower right corner of each panel. }
    \label{fig_app1_spindex}
    \end{figure}
}}
\end{center}

\setcounter{figure}{0} 
\begin{center}
\setlength{\tabcolsep}{1.2mm}{
{
    \begin{figure}
    \includegraphics[width=1.\linewidth]{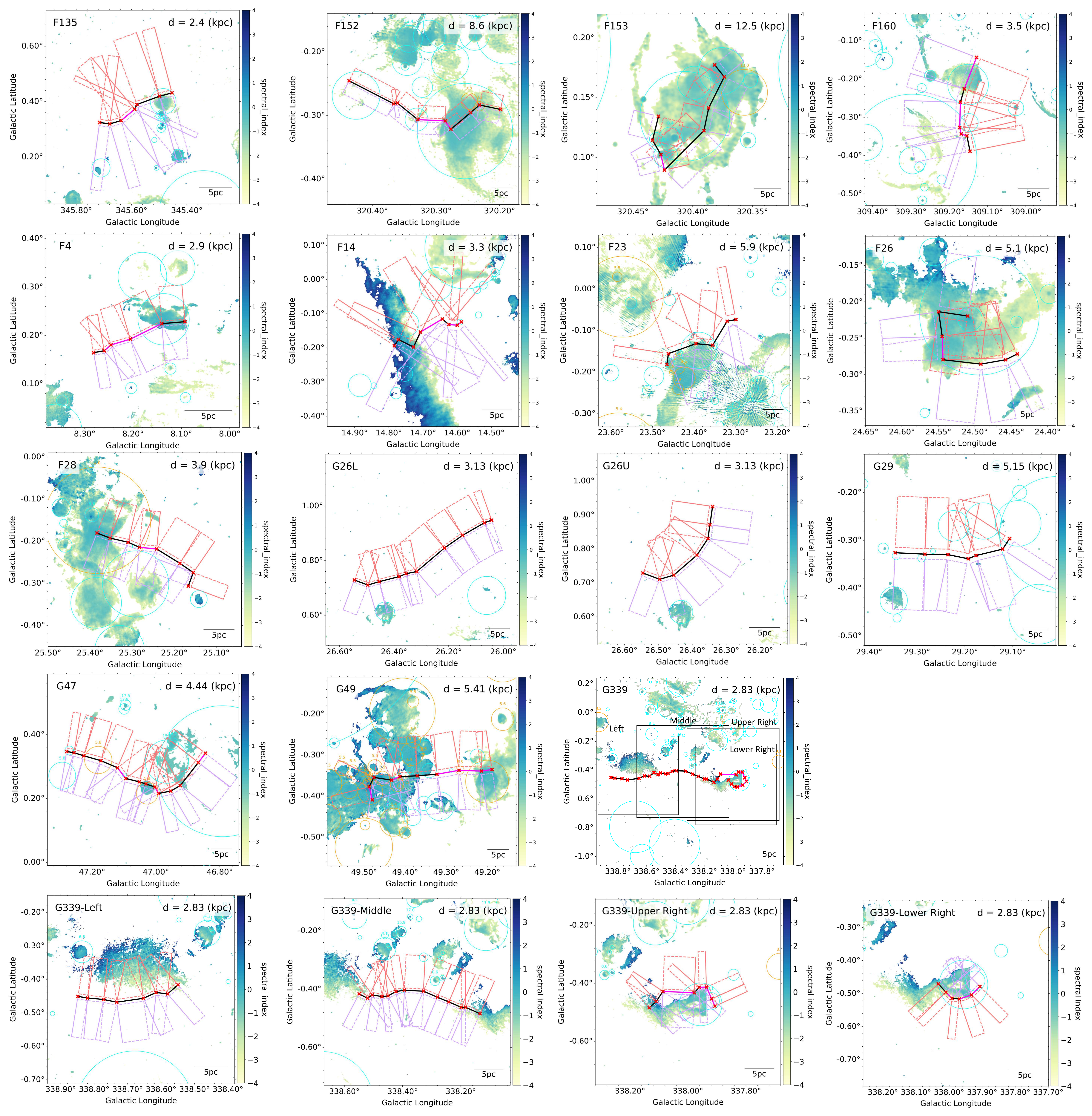}
    \caption{(continued)}
    \end{figure}
}}
\end{center}

\section{Calculation of uncertainties}\label{App_ErrorProp}
\subsection{Calculating the uncertainty maps of the two-dimensional Hi-GAL $\rm N(H_2)$ and $\rm T_d$ maps}
Based on the description in \citet{Marsh2017}, we write the expressions for the two-dimensional temperature-integrated $\rm N(H_2)$ maps and density-weighted mean $\rm T_d$ maps as follows:
\begin{equation}
    \label{formN}
    \rm N(H_2) = \sum_{i=1}^{12}N(H_2)_{T_i}
\end{equation}
\begin{equation} 
    \label{formT}
    \rm T_d=\frac{\sum_{i=1}^{12}T_i\times N(H_2)_{T_i}}{\sum_{i=1}^{12}N(H_2)_{T_i}}
\end{equation}
where $i$ represents the 12 temperature slices evenly spaced in logarithmic scale between 0 K and 50 K. $\rm N(H_2)_{T_i}$ denotes the column density in the $i$-th temperature slice, and $\rm T_i$ is the corresponding dust temperature of the slice. According to the law of error propagation, the uncertainties of the two-dimensional temperature-integrated $\rm N(H_2)$ and density-weighted mean $\rm T_d$ can be expressed as below:
\begin{equation} 
    \label{formsigN}
    \rm \sigma_{N(H_2)}=\sqrt{\sum_{i=1}^{12}\sigma^2_{N(H_2)_{T_i}}}
\end{equation}
\begin{equation}
    \label{formsigT}
    \rm \sigma_{\rm T_d}=\sqrt{
    \sum_{i=1}^{12}\left[\frac{T_i(\sum_{j=1}^{12} N(H_2)_{T_j}-N(H_2)_{T_i})-T_d\sum_{j=1}^{12}N(H_2)_{T_j}}{\left[\sum_{j=1}^{12} N(H_2)_{T_j}\right]^2}\right]^2\sigma^2_{N(H_2)_{T_i}}}
\end{equation}
where $\sigma_{\rm N(H_2)_{T_i}}$ represents the uncertainty of the column density in the $i$-th temperature slice, can be accessed from the corresponding uncertainty data cubes provided by \citet{Marsh2017}. Using the two expressions above, we computed the two-dimensional uncertainty maps of the two-dimensional temperature-integrated $\rm N(H_2)$ and density-weighted mean $\rm T_d$ maps provided by \citet{Marsh2017}. Based on the radial profiles and their corresponding uncertainty profiles, the average uncertainties of the $\rm N(H_2)$ and $\rm T_d$ radial profiles for the 35 LSFs are approximately 10.6\% and 7.9\% of the central peak of the $\rm N(H_2)$ radial profile, respectively, and their magnitudes remain relatively stable across different radii.

\subsection{Calculating the uncertainties of the $\alpha_{\rm asy}$ for each segment}
We first describe the procedure for calculating the uncertainty of $\alpha_{\rm asy}$, which is the parameter used to determine whether a radial profile of a single segment exhibits asymmetry. Since the mathematical forms of $\alpha_{\rm asy,N}$ and $\alpha_{\rm asy,T}$ are similar, we illustrate the process using $\alpha_{\rm asy,N}$ as an example. As defined in Equation \ref{eq_alpha}, $\alpha_{\rm asy,N}$ is derived from the radial profile extracted within the projection region, which uniformly take sample at fixed intervals corresponding to the pixel size. The calculation of $\alpha_{\rm asy,N}$ thus can be treated as summing up the total trapezoidal area under the $\rm N(H_2)$ radial profile:
\begin{equation}
    \alpha_{\rm asy,N}=\frac{\sum_{r=0.5w}^{n\times w-\Delta r}{(\rm N_{Side1}(r)+N_{Side1}(r+\Delta r))\Delta r/2}}{\sum_{r=0.5w}^{n\times w-\Delta r}{(\rm N_{Side2}(r)+N_{Side2}(r+\Delta r))\Delta r/2}}\equiv \frac{S_{\rm N1}}{S_{\rm N2}}
\end{equation}
we define ${S_{\rm N1}}$ and ${S_{\rm N2}}$ as the equation above, thus the uncertainty of $\alpha_{\rm asy,N}$ can be expressed as:
\begin{equation}
    \label{sig_alphaN}
    \begin{split}
        &\sigma_{\alpha_{\rm asy, N}}=\frac{S_{\rm N1}}{S_{\rm N2}}\times\sqrt{(\frac{\sigma_{S_{\rm N1}}}{S_{\rm N1}})^2+(\frac{\sigma_{S_{\rm N2}}}{S_{\rm N2}})^2},\\
        &\sigma_{S_{\rm N1}}=\frac{\Delta r}{2}\times\sqrt{\sum_{r=0.5w}^{n\times w-\Delta r}[\sigma_{\rm N_{\rm Side 1}(r)}^2+\sigma_{\rm N_{\rm Side 1}(r+\Delta r)}^2]},\\
        &\sigma_{S_{\rm N2}}=\frac{\Delta r}{2}\times\sqrt{\sum_{r=0.5w}^{n\times w-\Delta r}[\sigma_{\rm N_{\rm Side 2}(r)}^2+\sigma_{\rm N_{\rm Side 2}(r+\Delta r)}^2]},
\end{split}
\end{equation}
where $w$ is the width of the LSF, $n\times w$ is the upper limit of the radial profile sampling region, $\Delta r$ is the fixed length of sampling interval for each LSF. $\sigma_{\rm N_{Side 1}(r)}$ and $\sigma_{\rm N_{Side 2}(r)}$ denote the uncertainty of the temperature-integrated column density at radius $r$ on Side 1, 2 of the LSF, respectively. Their values are calculated using Equation \ref{formsigN}. For the $\rm T_d$ case, the uncertainty can be expressed in the same way:
\begin{equation}
    \label{sig_alphaT}
    \begin{split}
        &\alpha_{\rm asy,T}=\frac{\sum_{r=0.5w}^{n\times w-\Delta r}{(\rm T_{Side1}(r)+T_{Side1}(r+\Delta r))\Delta r/2}}{\sum_{r=0.5w}^{n\times w-\Delta r}{(\rm T_{Side2}(r)+T_{Side2}(r+\Delta r))\Delta r/2}}\equiv \frac{S_{\rm T1}}{S_{\rm T2}},\\
        &\sigma_{\alpha_{\rm asy, T}}=\frac{S_{\rm T1}}{S_{\rm T2}}\times\sqrt{(\frac{\sigma_{S_{\rm T1}}}{S_{\rm T1}})^2+(\frac{\sigma_{S_{\rm T2}}}{S_{\rm T2}})^2},\\
        &\sigma_{S_{\rm T1}}=\frac{\Delta r}{2}\times\sqrt{\sum_{r=0.5w}^{n\times w-\Delta r}[\sigma_{\rm T_{\rm Side 1}(r)}^2+\sigma_{\rm T_{\rm Side 1}(r+\Delta r)}^2]},\\
        &\sigma_{S_{\rm T2}}=\frac{\Delta r}{2}\times\sqrt{\sum_{r=0.5w}^{n\times w-\Delta r}[\sigma_{\rm T_{\rm Side 2}(r)}^2+\sigma_{\rm T_{\rm Side 2}(r+\Delta r)}^2]}
    \end{split}
\end{equation}
where $\sigma_{\rm T_{\rm Side 1}(r)}$ and $\sigma_{\rm T_{\rm Side 2}(r)}$ are the uncertainty of the density-weighted mean $\rm T_d$ at radius $r$ on Side 1, 2 of the LSF, calculated using Equation \ref{formsigT}. 

Here we describe the expected value of $\alpha_{\rm asy}$ for a symmetric radial profile with an uncertainty level comparable to that of the 35 LSFs. First, we simulated perfectly symmetric $\rm N(H_2)$ and $\rm T_d$ radial profiles at each LSF distance by smoothing the RCW 120 filament’s radial profile to the corresponding resolution. For the $\rm N(H_2)$ profile, since its shape is assumed to approximate a Plummer-like profile with $p=4$, we adopted the Plummer $p=4$ fit to the $\rm N(H_2)$ radial profile on the side of the RCW 120 filament farther from the central H\textsc{ii} region to construct the symmetric $\rm N(H_2)$ profile. For the $\rm T_d$ profile, because its functional form cannot be reliably associated with any known analytic expression, we directly mirrored the $\rm T_d$ radial profile on the side of the filament far from the central region to obtain a fully symmetric simulated $\rm T_d$ profile.

Next, using the $\rm N(H_2)$ and $\rm T_d$ radial profiles and their corresponding uncertainty profiles for the 35 LSFs, we determined average uncertainties of $\sigma_{\rm N}=0.106\times{\rm N(H_2)_{max}}$ and $\sigma_{\rm T}=0.079\times{\rm T_{d,max}}$, where ${\rm N(H_2)_{max}}$ and ${\rm T_{d,max}}$ are the maximum density and temperature values within the segment-length–weighted average radial profiles of each LSF. Both uncertainties remain relatively stable across different sampling radii.

\begin{figure*}[!t]
    \centering
    \includegraphics[width = 1.\textwidth]{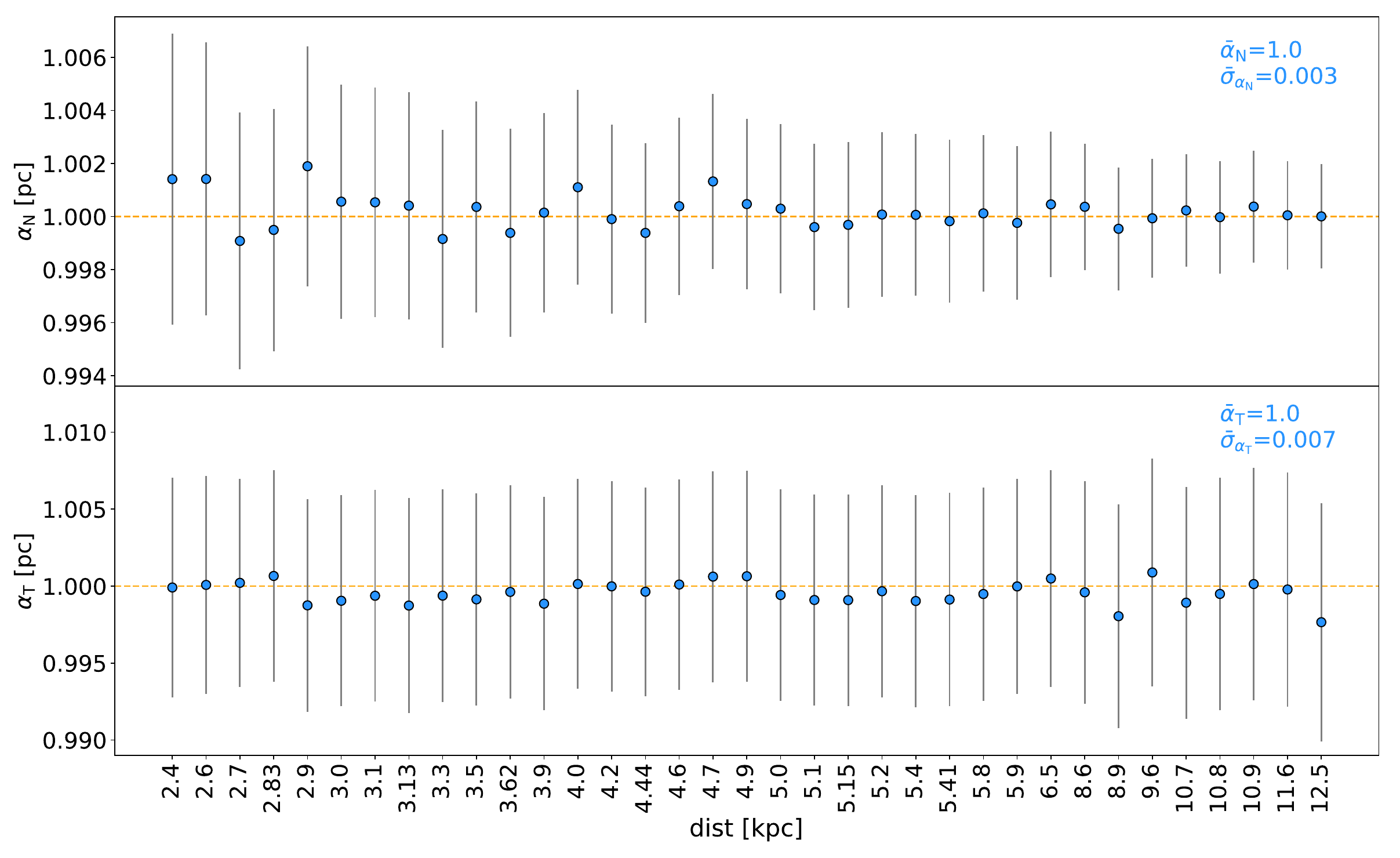}
    \caption{The averaged $\alpha_{\rm N}$ and $\alpha_{\rm T}$ values of the 100 simulated symmetric radial profiles with noise at each distance of the 35 LSFs. The orange horizontal dashed line marks the perfectly symmetric case with $\alpha_{\rm asy}=1$. All simulated radial profiles have $\alpha_{\rm N}$ and $\alpha_{\rm T}$ consistent with 1 within the $1\sigma$ range, and are therefore classified as symmetric. The blue annotation in the upper-right corner of each panel indicates the averaged $\alpha_{\rm asy}$ and the corresponding $\sigma_{\alpha}$ over all distances.}
    \label{fig_app2_simsymprof}
\end{figure*}

We assume that the uncertainty at each radius of the symmetric radial profile follows a uniform distribution within the range $\pm\sigma$. For each LSF distance, we generated 100 realizations of uncertainty profiles satisfying this assumption and added them to the simulated symmetric radial profile. This yielded 100 noise-added symmetric radial profiles per distance. We then computed $\alpha_{\rm asy}$ and its uncertainty for each simulated profile following the procedure described in Section \ref{subsec:asy_param}. The resulting values are shown in Figure \ref{fig_app2_simsymprof}. At all distances, the simulated symmetric profiles have $\alpha_{\rm N}$ and $\alpha_{\rm T}$ values whose uncertainty ranges overlap with unity, and thus are still identified as symmetric. This confirms that our use of $\alpha_{\rm asy}$ as a diagnostic for detecting asymmetry in radial profiles is robust.

\subsection{Calculating the uncertainty of the $|1-\alpha_{\rm asy}|$ for each LSF}
Since our goal is to use the segment-length–weighted average of $|1-\alpha_{\rm asy}|$ to represent the overall degree of asymmetry in the surroundings across the two sides of an LSF, the error bars of $|1-\alpha_{\rm asy}|$ shown in Figure \ref{fig4_alpha} include not only the mean uncertainty of $\alpha_{\rm asy}$ for all segments, but also the weighted sample variance introduced when computing the segment-length–weighted average of $|1-\alpha_{\rm asy}|$ for the segments within the same LSF:
\begin{equation}
    \begin{split}
        &\sigma_{|1-\alpha_{\rm asy}|}=\sqrt{\sigma_{\bar{\alpha}_{\rm asy}}^2+s^2},\\
        &\sigma_{\bar{\alpha}_{\rm asy}} = \sqrt{\sum_iw_i^2\sigma_{a_{{\rm asy},i}}^2},\\
        &s = \sqrt{\frac{\sum_iw_i((1-\alpha_{\rm asy})-(1-\bar{\alpha}_{\rm asy}))^2}{\sum_iw_i}}
    \end{split}
\end{equation}
where $w_i$ is the length of the $i$-th segment, and $\sigma_{a_{{\rm asy},i}}$ is the uncertainty of $\alpha_{\rm asy}$ for that segment. Across our sample of 35 LSFs, the averaged ratio between $\sigma_{\bar{\alpha}_{\rm asy}}$ and $s$ is 0.11 and 0.46 for the $\rm N(H_2)$ and the $\rm T_d$ case, respectively. These ratios demonstrates that the LSFs in our sample truly reside in variable column-density and relatively stable temperature environments.

\bibliography{reference}{}
\bibliographystyle{aasjournal}

\end{document}